\documentclass{ws-ijbc}
\usepackage{ws-rotating} 
\begin{document}
       
\catchline{}{}{}{}{} 

\markboth{Katsanikas et al}{Instabilities  and  stickiness in a 3D rotating 
galactic potential}

\title{Instabilities  and  stickiness in a 3D rotating galactic potential}
\author{M. KATSANIKAS}
\address{Research Center for Astronomy, Academy of Athens\\
  Soranou Efessiou 4,  GR-11527 Athens, Greece}
\address{Section of Astrophysics, Astronomy and Mechanics, \\Department of
  Physics, University of Athens, Greece\\ mkatsan@academyofathens.gr}
\author{P.A. PATSIS and G. CONTOPOULOS}
\address{Research Center for Astronomy, Academy of Athens\\
  Soranou Efessiou 4,  GR-11527 Athens, Greece\\
   patsis@academyofathens.gr, gcontop@academyofathens.gr}

\maketitle

\begin{history}
\received{(to be inserted by publisher)}
\end{history}

\begin{abstract}
We study  the  dynamics in the neighborhood of simple and double unstable 
periodic orbits in a rotating 3D autonomous Hamiltonian system of galactic type. In 
order to visualize the four dimensional spaces of section we use the method of 
color and rotation. We investigate the structure of the invariant manifolds 
that we found in the neighborhood of simple and double unstable periodic orbits in 
the 4D spaces of section. We consider orbits in the neighborhood of the families x1v2, belonging to the x1 tree, and the z-axis (the rotational axis of our system).
Close to the transition points from stability to simple instability, in the neighborhood of the bifurcated simple unstable x1v2 periodic orbits we encounter the phenomenon of stickiness as the asymptotic curves of the unstable manifold surround regions of the phase space occupied by rotational tori existing in the region. For larger energies, away from the bifurcating point, the consequents of the chaotic orbits form clouds of points with mixing of color in their 4D representations. In the case of double instability, close to x1v2 orbits, we find clouds of points in the four dimensional spaces of section. However, in some cases of double unstable periodic orbits belonging to the z-axis family we can visualize the associated unstable eigensurface. Chaotic orbits close to the periodic orbit remain sticky to this surface for long times (of the order of a Hubble time or more). Among the orbits we studied we found those close to the double unstable orbits of the x1v2 family having the largest diffusion speed. The sticky chaotic orbits close to the bifurcation point of the simple unstable x1v2 orbit and close to the double unstable z-axis orbit we have examined, have comparable diffusion speeds. These speeds are much slower than of the orbits in the neighborhood of x1v2 simple unstable periodic orbits away from the bifurcating point, or of the double unstable orbits of the same family having very different eigenvalues along the corresponding unstable eigendirections.
\end{abstract}


\section{Introduction}

The aim of this paper is to study the dynamics in the neighborhood of unstable
periodic orbits in a rotating three dimensional (hereafter 3D) autonomous 
Hamiltonian system of galactic type. Instability of periodic orbits in such a 
dynamical system is of three kinds: simple, double and 
complex instability. The case of complex instability was studied in a recent 
paper by Katsanikas et al (Katsanikas et al. 2011). Here we study the two 
other cases. Furthermore, we examine the stickiness effects in our 3D system. This is the 
first time that the stickiness phenomenon is \textit{explicitly} investigated in association with
these kinds of instability (simple and double) in a 3D autonomous Hamiltonian system. For this purpose we consider the intersections of the orbits by a space of section. In 
a 3D autonomous Hamiltonian system the space of section has 4 dimensions. So, for 
the visualization of the 4D spaces of section it is convenient to use the
method of color and rotation (Patsis \& Zachilas 1994). Rotation is used for 
understanding the geometry of a 3D projection and color is used to visualize 
the 4th dimension.

3D rotating autonomous Hamiltonian systems are of special importance in
Galactic Dynamics. They represent in a realistic way rotating galaxies, in
which rotation plays a key role in the dynamics on the equatorial plane, as
well as in the third dimension. The most important family for the vertical 
morphology of rotating disk galaxies is a family called x1v1 by 
(Patsis et al. 2002). 
The x1v1 family is introduced at the transition 
from stability to simple instability ($S\rightarrow U$) as the stability of the central,
planar family x1 undergoes a $S \rightarrow U \rightarrow S$ transition within a small 
Jacobi integral interval at the vertical 2:1 resonance region (for a
definition of resonances in rotating galactic potentials see Contopoulos
 pages 379 and 422). At 
the $U \rightarrow S$ transition, another family of periodic orbits (hereafter
p.o.), x1v2, is bifurcated from x1 as simple unstable (Skokos et al. 2002). For larger Jacobi constants x1v2 becomes double unstable. Thus, it possesses both kinds of instabilities we want to study. In this paper we focus on x1v2 and we examine the structure of the  phase space in its neighborhood, when it is simple or double unstable. However, in order to investigate in depth the  dynamics in the neighborhood of double unstable periodic orbits we will also study a case referring to z-axis orbits.

The importance of the dynamical phenomena at the 2:1 vertical resonance is due
to the role of the associated families in shaping the vertical morphology of
the 3D disks. They form to a large extent the peanut-shaped structure, which
is characteristic of many disk galaxies observed edge-on (Patsis et al 2002). 
In addition, chaotic motion in the neighborhood of unstable p.o.'s has been
proven to be significant for the dynamics of rotating galactic disks
\textit{on} their equatorial planes (for a review see e.g. Contopoulos 2009). 
In that respect the knowledge of the structure of phase space in the neighborhood of 3-dimensional p.o. in galactic systems is the first step for the investigation of a possible contribution of chaotic orbits to the observed vertical profiles of this type of galaxies.

The space of section that we use in this paper for investigating the orbital behavior close to x1v2 orbits is defined by  $y=0$  and we 
consider intersections with  $\dot y>0$. Hereafter by ``intersections'' we will refer to intersections, which fulfill these properties, until we start discussing the dynamics close to z-axis. Then we will change space of section.  Every  consequent in our cross 
section is  determined by 4 Cartesian coordinates, e.g. $(x,\dot x,z,\dot z)$. In 
the method of color and rotation we project the consequents in a 3D 
subspace and observe them from different viewpoints. In the text the viewpoints are given 
in spherical coordinates. 
The unit  for the distance $d$ of the
consequents of the surface of section from the observer corresponds to the longest
side of the  bounding box of the figure (more details can be found in 
Katsanikas and Patsis 2010).
In practice we rotate the 3D figure in the computer 
screen in order to find the most appropriate viewpoint for the presentation on  
the 2D surface of the pages of a paper. We select the viewpoint that represents best the morphology of the 
structures  formed  by the consequents in the 3D projections of the 4D space of 
section in order to facilitate the reader to follow our description. Then we color every point according the value of its  4th
coordinate. For this reason  we  use  the ``Mathematica'' package and we map  
the interval of the values of the fourth coordinate  
into $[0,1]$. 
\begin{itemize}
\item A well defined structure in all 3D projections and a smooth color variation along them is evidence for regular motion.
\item Mixing of colors indicates
chaotic behavior of the consequents in 4D (Patsis \& Zachilas 1994, Katsanikas
\& Patsis 2010)
\end{itemize}

\subsection{Stability of periodic orbits}

The linear stability of the periodic orbits is studied, following the
analysis  by Broucke (Broucke 1969)  in the case of the elliptic restricted
three body problem and the method used by Hadjidemetriou (Hadjidemetriou 1975) 
in the 
three body problem. 
The method has been also described in a recent earlier paper of us in the
series of papers studying the dynamics in the neighborhood of p.o. in 3D
rotating galactic potentials (Katsanikas \& Patsis 2010). 
The characteristic equation in a 3D autonomous hamiltonian system is of the form $\lambda^4 + a \lambda^3 + \beta \lambda^2 + a \lambda +1 = 0 $. and has 4 eigenvalues $\lambda_{1}$, $1/\lambda_{1}$, $\lambda_{2}$, $1/\lambda_{2}$, where $\lambda_i, \frac{1}{\lambda_i} = \frac {- b_i \pm \sqrt{b_i^2 - 4}}{2}$, $i=1,2$, with $b_{1,2}=\frac{a \pm \sqrt{\Delta}}{2}$ and $\Delta = a^2 - 4 (\beta - 2)$. Thus, the stability of the p.o. depends on the two stability indices $b_1$ and $b_2$ and on the quantity $\Delta$.

For  $\Delta > 0,\; |b_1| < 2$  and  $|b_2| < 2$, all four eigenvalues are on the unit circle in the complex plane and the orbit is called
stable $(S)$ (see e.g. Contopoulos 2002, Contopoulos \& Magnenat 1985).

Then we have three kinds of instability:
\begin{enumerate}

\item simple instability $(U)$: 
$\Delta > 0$  and $|b_1| > 2, \; |b_2| < 2$ or $|b_1| < 2,\;|b_2| > 2$, (two eigenvalues on the unit circle and two on the real axis)

\item double instability $(DU$):
$\varDelta > 0$  and $|b_1| > 2$  and  $|b_2| > 2$ (four eigenvalues on the real axis)

\item complex instability ($\varDelta$) (all four complex eigenvalues are off the unit circle in the complex plane):
 $\varDelta< 0$
\end{enumerate}
(for details see Contopoulos 2002, p.286). Hereafter we will indicate the periodic orbits as $S$, $U$, $DU$ and $\varDelta $ according to their stability.
A generalization of the three kinds of instability  in Hamiltonian systems of $N>3$
degrees of freedom can be found in  Skokos (Skokos 2001).

In section 2 we present our Hamiltonian  system. In section 3 
we discuss the stability of a family of p.o. we call x1v2. In section 4 we study the orbital behavior close to simple unstable orbits of x1v2, in section 5 we 
investigate  the dynamical behavior in the neighborhood of double unstable orbits of x1v2, as well as of the z-axis families, in section 6 we compare the diffusion of the chaotic orbits away from the p.o. and finally in section 7 we discuss and enumerate our conclusions.
 
\section{The Hamiltonian System}

The Hamiltonian of our system in Cartesian coordinates is:

\begin{equation}
H(x, y, z, \dot x, \dot y, \dot z)=
\frac {1}{2}(\dot x^2 + \dot y^2 + \dot z^2) + \Phi(x, y, z)
- \frac{1}{2} \Omega_b^2 (x^2 + y^2),
\end{equation}

where  $\Phi(x,y,z)$ is the potential:

\begin{equation}
\Phi(x,y,z)= 
-\frac{GM_{1}}{(x^2+ \frac{y^2}{q_a^2} +[a_{1}+(\frac{z^2}{q_b^2}+
b_{1}^2)^{1/2}]^2)^{1/2}}-
 \frac{GM_{2}}{(x^2+ \frac{y^2}{q_a^2} +[a_{2}+(\frac{z^2}{q_b^2}+b_{2}^2)
^{1/2}]^2)^{1/2}}
\end{equation}

This potential in its axisymmetric form $(q_a=1,q_b=1)$ can be considered as
an approximation of the potential of the Milky Way (Miyamoto \& Nagai 1975). 

In our units, the distance $R$=1 corresponds to 1~kpc. The velocity unit
corresponds to 209.64~$km/sec$. The Jacobi constant  $Ej=1$ corresponds to 
43950 $(km/sec)^2$ and we will call it ``energy''. For the parameters we have 
used the following values: $ a_{1}=0 \; kpc,\; b_{1}=0.495
\; kpc,\; M_{1}=2.05 \times 10^{10} \; M_{\odot},\; a_{2}=7.258 \;
kpc,\; b_{2}=0.520 \; kpc,\; M_{2}=25.47 \times 10^{10} \;
M_{\odot},\; q_a = 1.2,\; q_b = 0.9$. For the angular velocity we used the
value $\Omega_b=0.045$ in our units. The parameters $q_a, q_b$ determine the 
geometry of the disks, while $a, b$ are scaling factors. The specific values
of the geometrical parameters do not have a particular physical meaning. The
dynamical phenomena we study in our paper are typical for rotating triaxial
systems. The $\Omega_b$ value we use here is different from the one we used in
(Katsanikas \& Patsis 2010) in order to have larger regions in the parameters' space with non-escape orbits close to double unstable periodic orbits (Section 5).

\section{Stability Diagram}
\begin{figure}[t]
\begin{center}
\begin{tabular}{cc}
\resizebox{95mm}{!}{\includegraphics{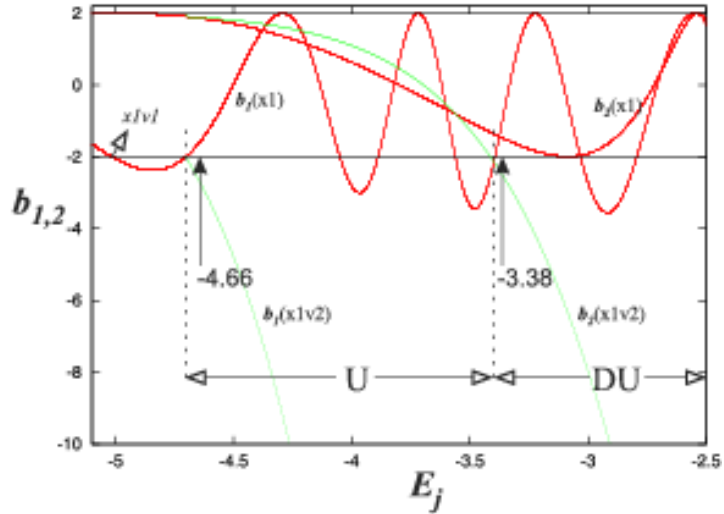}}\\
\end{tabular}
\caption{The stability diagram for  $-5 < E_j < -2.3$, that shows the stability
  indices of the families x1 (red line) and x1v2 (green line).}
\label{dsimple1}
\end{center}
\end{figure}

 In order to study the evolution  of the stability of a family of periodic 
orbits as a parameter of our system varies we use the stability diagram 
(Contopoulos 2002 p.287). The stability diagram gives the values of
 the coefficients $b_1$ and $b_2$ for successive periodic orbits in a family 
 by changing only one parameter of the system (in our case $E_j$), while keeping the other parameters constant. In Fig.~\ref{dsimple1} we observe in such a diagram how  the family of periodic 
orbits x1v2 (Skokos et al. 2002) bifurcates from the 2D central family x1 on
the plane $z=0$ at the $U\rightarrow S$ transition that happens in the
vertical 2:1 resonance region (Contopoulos 2002  p.390). This bifurcation   occurs for 
$E_j\simeq-4.7$. We observe that the stability parameters $b_1$, $b_2$ of the 
family x1v2 decrease  as $E_j$ increases. The  value of $b_2$  is 
initially about $1.95$ at $E_j=-4.7$ (left side of diagram). When 
$b_2$  becomes $-2$ we have a  transition of x1v2 from simple to double 
instability ($U \rightarrow DU$) (for $E_j \simeq -3.41$) and the x1v2
periodic orbits  become  double unstable. The $E_j$ interval of x1v2, where it is  
simple unstable ($-4.7 < E_j < -3.41$) and the $E_j$ interval of x1v2 where it is 
double unstable ($-3.41 < E_j < -2.3$), will be called hereafter
regions U and DU respectively (Fig.~\ref{dsimple1}).

\section{Simple Instability}
First we study the orbital behavior of the x1v2 family in the region U. In the following applications 
we use the set of projections $(x,\dot x,z)$  and $(\dot x,z,\dot z)$. However, we 
find qualitatively similar results by using the $(x,\dot x,\dot z)$  and $(x,z,\dot z)$ 
projections respectively. The use of the two projections gives us all needed information to understand the structure of the distribution of the consequents in the 4D space by means of the color-rotation method, as we will see below.

The initial conditions to be studied are chosen in two different ways. First we  investigate the dynamical behavior of the orbits in 
the neighborhood of the  simple unstable periodic orbit x1v2 at
$E_j=-4.66$ by varying the  values and the direction 
($\Delta x$, $\Delta \dot x$, $\Delta z$ or $\Delta \dot z$) of the 
perturbation. Secondly, we consider a constant perturbation 
$\Delta x =10^{-4}$ and we apply this perturbation to the initial conditions 
of several periodic  orbits of the x1v2 family in the region U for different 
$E_j$ values. We start by keeping $E_j$ constant and investigate as a sample case the dynamical behavior very close to the point where we have the $U\rightarrow S$ transition of the x1 family and the bifurcation of the family x1v2 as $U$.

\subsection{Orbital behavior close to the bifurcating point - A sample case}
\begin{figure}[t]
\begin{center}
\begin{tabular}{cc}
\resizebox{96mm}{!}{\includegraphics{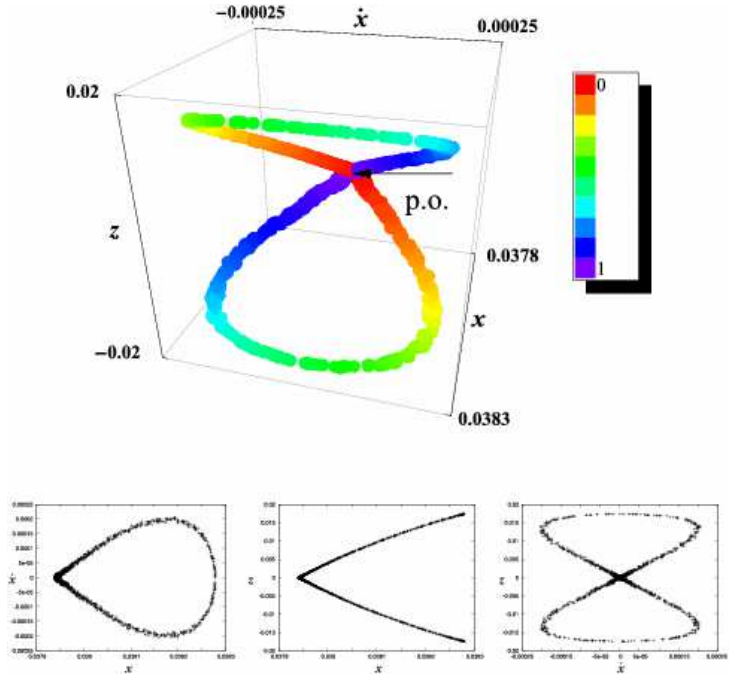}}\\
\end{tabular}
\caption{Upper part: A 4D representation of the orbit ``u'' for 2580 consequents. The consequents
are given in the $(x,\dot x,z)$ projection, while every point is colored 
according its $\dot z$ value. We observe that an double loop ribbon is
formed with a smooth color succession on it (cf. with the color bar on the 
right). The location of the periodic orbit (p.o.) is indicated by an arrow. Our 
point of view in  spherical  coordinates is $(\theta, \phi) = ( 30^{o}, 
108^{o})$. Lower part: The three 2D projections $(x,\dot x,), (x,z)$ and $(\dot x,z)$. The ranges on the axes are as in the main figure.}
\label{dsimple2}
\end{center}
\end{figure}

The energy
$E_j=-4.66$  is indicated with an arrow in Fig.~\ref{dsimple1}. The  periodic orbit has initial conditions 
$(x_0,\dot x_0,z_0,\dot z_0)=(0.037941612,0,0,0.42902353)$. In order to 
study the orbital behavior in its neighborhood we added a small perturbation in the 
$\dot z$-direction equal to $\Delta \dot z=-6\times 10^{-6}$. We  call this perturbed 
orbit ``u''. The first 2580 consequents of orbit ``u'' define a ribbon of the 
form of a double loop in the $(x,\dot x,z)$ subspace (Fig.~\ref{dsimple2}).
In the bottom of the figure we give also from left to right the 2D projections $(x,\dot x),(x,z)$ and $(\dot x,z)$ in order to help the reader understand the topology of the structure. The two lobes of the double loop are inclined with respect to each other and form an angle around $50^{o}$ in the $(x,z)$ projection (middle panel in the lower part of Fig.~\ref{dsimple2}). The $(x,z)$ projection clearly also shows that the thickness of the lobes is very small (but not zero)  and this is the reason we describe them as ribbons. Essentially the branches of the lobes are thin tubes. 
The 4D representation of this structure, colored according to the normalized $\dot z$ values and using $(x,\dot x,z)$ for the 3D spatial projection is given in Fig.~\ref{dsimple2} (top). For the 3D projection we use here a point of view defined by the angles $(\theta, \phi) = (30^{o}, 108^{o})$.
We observe that this ribbon has a smooth color variation from  red to violet, in agreement with structures found in the neighborhood of simple periodic $(U)$ orbits by Patsis and Zachilas (1994). The structure of the phase space 
close to simple unstable periodic orbits has been studied for the first time  by 
Magnenat (1982). However, Magnenat used 2D projections and these did not allow 
the full understanding of the 4D structure of the double loop objects. In 
the present case, our 4D representation allows us to understand the details of this structure.  

By inspection of Fig.~\ref{dsimple2} we infer that the color variation along the two 
branches is such that at the intersection region we have the meeting of two different shades.
This means that 
in this region the points have  different values in the 4th dimension and 
there is no real intersection in the 4D space of section. Consequently  the 
ribbon-like double loop structure has not an intersection in 4D. 
We observe that the colors at the region of the intersection are not only different, but that they are at the two extremes of our color scale, i.e. violet and red (see right side of Fig.~\ref{dsimple2}). The colors in this region are clustered around two values $(\dot z = \pm 0.429)$ very close to the $\dot z$ initial condition of the periodic orbit $\dot z = 0.42902353$. 

Figure \ref{dsimple2} reflects the symmetry of the galactic potential we study. Due to this symmetry, the families of p.o. that bifurcate at the vertical resonances from the planar x1 family come in pairs with mirror morphology  with respect to the equatorial plane of the system in the configuration space, i.e. x1v1 and x1v1$^{\prime}$, x1v2 and x1v2$^{\prime}$ etc. The p.o. of x1v2 and x1v2$^{\prime}$ at every $E_j$ differ only in the sign of their $\dot z_0$ initial conditions. In the 3D $(x,\dot x,z)$ projection we use in Fig.~\ref{dsimple2}, the initial conditions of x1v2 and x1v2$^{\prime}$ are identical and the p.o. is in the center of the intersection region of the two branches of the double loop ribbon (indicated with an arrow). 
The clustering of the colors around the two values $(\dot z_0 \approx \pm 0.429)$ in the central region is due to the coexistence of x1v2 and x1v2$^{\prime}$. The consequents drift away from the central region of the double loop by deviating from the two $(\pm \dot z_0)$ initial conditions with different colors. The first 2580 consequents are along the double loop structure, thus they indicate a regular rather than chaotic motion.

\begin{figure}[t]
\begin{center}
\begin{tabular}{cc}
\resizebox{90mm}{!}{\includegraphics{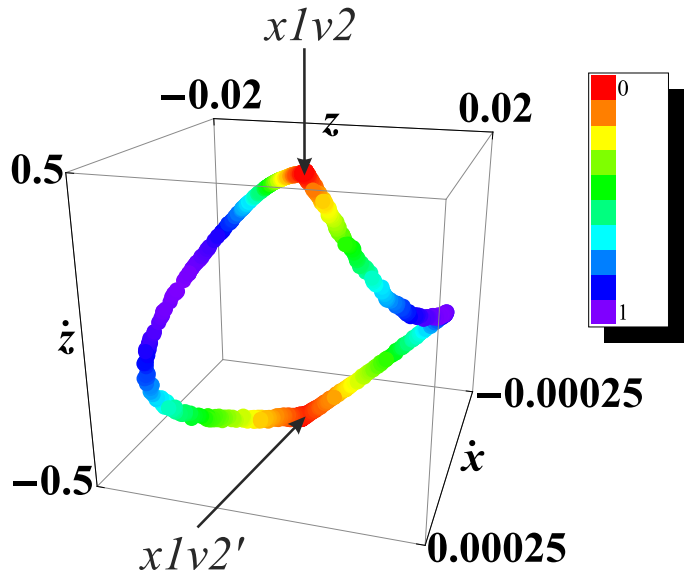}}\\
\end{tabular}
\caption{2580 consequents of the orbit ``u'' in the 3D subspace 
$(\dot x,z,\dot z)$. This time every point is colored according its 
$x$ value. The arrows  indicates the locations of the periodic orbits x1v2 and x1v2$^{\prime}$.
Our point of view in  spherical  coordinates is 
$(\theta, \phi) = (45^{o},144^{o})$. }
\label{dsimple3}
\end{center}
\end{figure}

We reach the same conclusion by using other 3D projections for the spatial coordinates of the consequents. 
In Fig.~\ref{dsimple3}  the first 2580  consequents of the orbit ``u'' in the 3D 
projection $(\dot x,z,\dot z)$
form  again a  ribbon-like surface, however this time with a single loop morphology. Color is determined by the $x$ value of the consequents and  we observe a smooth color variation
 along the ribbon. In this case the two representatives of the p.o x1v2 and x1v2$^{\prime}$ are at two different positions in the $(\dot x,z,\dot z)$ space, since they have different $\dot z_0$ initial conditions. Their positions are indicated with arrows in Fig.~\ref{dsimple3}. The regions close to both of them are colored red (same $x$ values) and departing from them,  red becomes orange, then yellow, green, light blue, blue and finally violet. The ribbon  is a 4D object. From both
Figs. \ref{dsimple2} and \ref{dsimple3} we conclude that the first 2580 
consequents form a well-defined ribbon-like surface that has no 
self-intersections in 4D. 

A similar analysis can be done by means of the two other 3D projections $(x, \dot x,\dot z)$ and $(x,z,\dot z)$, which give 4D structures morphologically similar to $(x,\dot x,z)$ and $(\dot x,z,\dot z)$ respectively. It is evident that the orbit ``u'' in the neighborhood of a $U$ x1v2 p.o. for 2580 consequents is confined in a particular region of the space of section with dimensions ($\Delta x \times \Delta \dot x \times \Delta z \times \Delta \dot z) < (0.00035 \times 0.0005 \times 0.04 \times 1) $. 
As we will see below, for larger integration times the orbit diffuses to a larger region of the phase space. 
The structures we find in  the phase space indicate a sticky behavior and this  will be further analyzed in section 4.4.  

We note that although in general we have different colors in the intersections of double loop type structures, like the one in Fig.~\ref{dsimple2}a, in exceptional cases we can have also real intersections in the 4D space. Such a case is encountered in the neighborhood of $U$ z-axis orbits in 4D spaces of section defined by $z=0$. The z-axis orbits lie entirely on the rotational, z axis, of the system. They have initial conditions $(x,\dot x, y,\dot y)=(0,0,0,0)$ and are discussed in detail in section 5.2 in the context of the study of double instability.

\begin{figure}[t]
\begin{center}
\begin{tabular}{cc}
\resizebox{90mm}{!}{\includegraphics{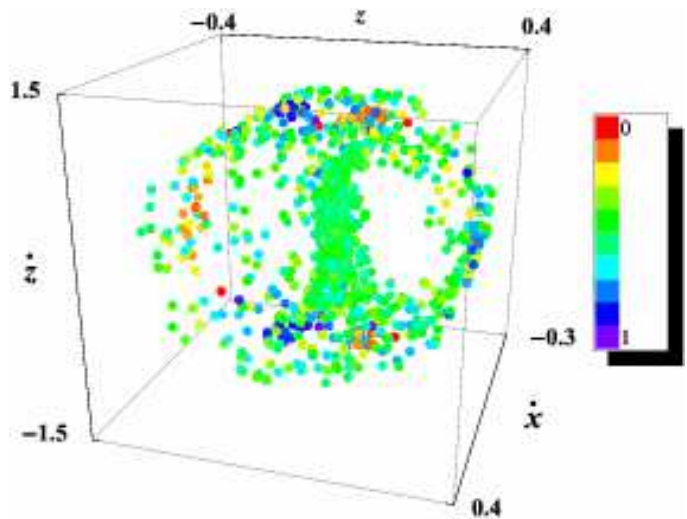}}\\
\end{tabular}
\caption{Mixing of colors for the first 1000 consequents of the orbit ``u'' (beyond the first 2580 consequents) that 
deviate from the 4D ribbon surface. The 3580 points totally are depicted in the 3D subspace 
$(\dot x,z,\dot z)$ and the color represents the 4th dimension $x$ of the 
points.  Our point of view in spherical  coordinates is 
$(\theta, \phi) = ( 45^{o},144^{o})$.}
\label{dsimple7}
\end{center}
\end{figure}

If we compute a larger number of intersections
we observe a drastic change in the topology of the ribbon-like surface. The 
points deviate from the well-defined 4D structures that we described above and form a cloud of points around them in the 3D projections with mixed colors in their 4D representations. However, in the particular case we study here, for about 3580 intersections, i.e. for 1000 consequents more than those building the structures in Figs.~\ref{dsimple2} and \ref{dsimple3}, in the 3D projections that include the $z$ and $\dot z$ coordinates we discern a \rotatebox[origin=c]{90}{$\Theta \:$} structure (i.e. a greek letter $\Theta$ rotated by $90^{o}$) surrounded by a ring. This can be seen in Fig.~\ref{dsimple7}, where we depict the consequents of the orbit ``u'' in a $(\dot x,z,\dot z)$ projection and color them according to their $x$ values. This new structure in the 3D projection of the phase space is maintained for several thousands of intersections more. We have again a confinement of the consequents in a restricted volume of the phase space, which is larger than previously. After 3580 intersections the consequents extend within a region ($\Delta x \times \Delta \dot x \times \Delta z \times \Delta \dot z) < (0.18 \times 0.7 \times 0.8 \times 2.4) $. In the new situation we have on one hand the appearance of structures in certain projections which indicates order, and on the other hand mixing of colors, which indicates chaoticity.

\begin{figure}
\begin{center}
\begin{tabular}{cc}
\resizebox{90mm}{!}{\includegraphics{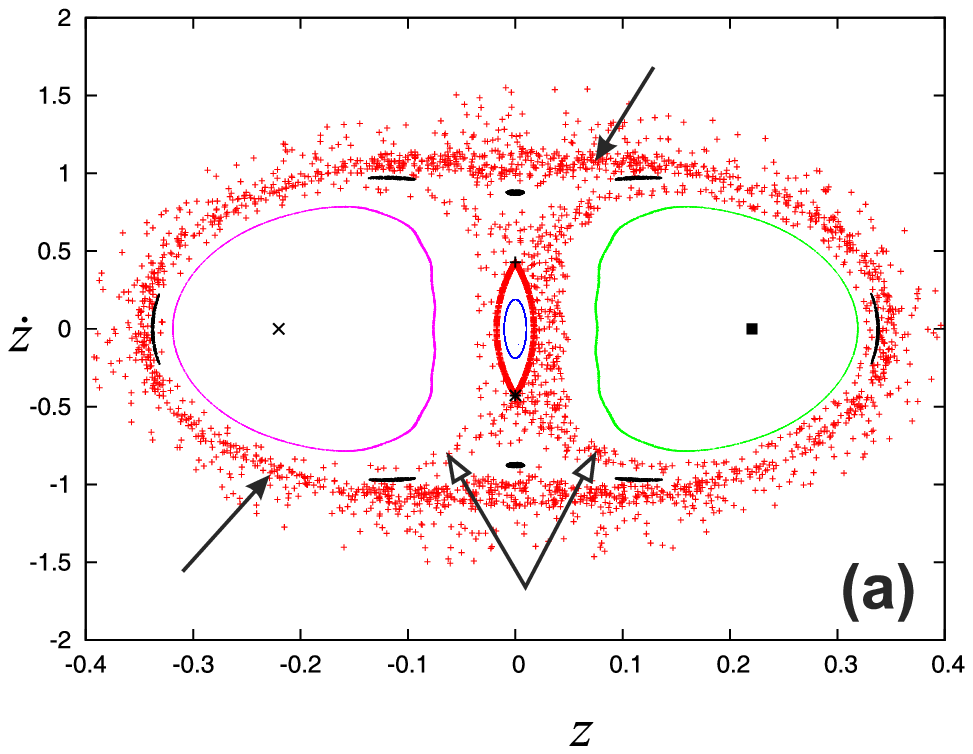}}\\ 
\resizebox{90mm}{!}{\includegraphics{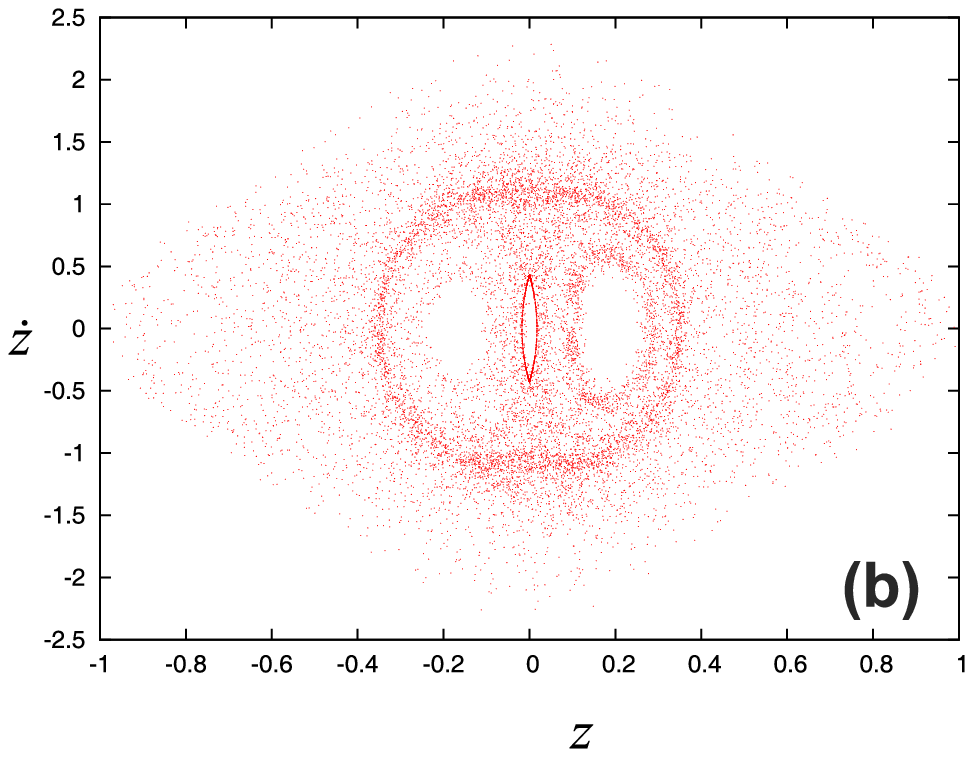}}\\
\end{tabular}
\caption{(a) 5000 consequents of the orbit ``u'' in the 2D $(z,\dot z)$ projection. The first 2580 consequents form the red ``ellipse with corners'' in the central part of the figure. The location of x1v2 and x1v2$^{\prime}$ are indicated with black symbols on it (up and down). The blue ellipse inside the red one is the projection of a rotational torus around x1. The rest of the consequents, after the first 2580, diffuse in phase space, but remain in the region around the rotational tori belonging to the stable p.o. of the x1v1 and x1v1$^{\prime}$ families. The \textit{projections} of the two tori around each of the two stable p.o. (indicated also with black symbols in the middle of the ``empty'' regions) appear in this case like green and magenta invariant curves. Arrows point to the lobes and ring structure (see text). (b) 15000 consequents of the orbit ``u'' in the 2D $(z,\dot z)$ projection. They visit all available regions of the phase space. However the sticky regions formed during the integration time are clearly discernible.}
\label{x1v1v2}
\end{center}
\end{figure}

\begin{figure}
\begin{center}
\begin{tabular}{cc}
\resizebox{70mm}{!}{\includegraphics{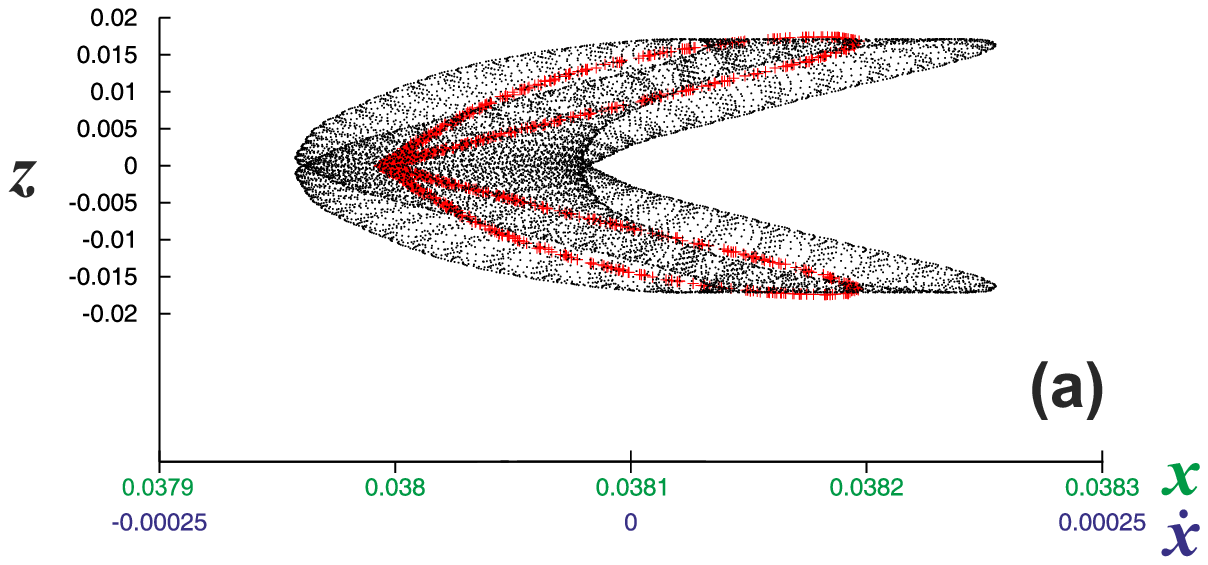}}\\
\resizebox{70mm}{!}{\includegraphics{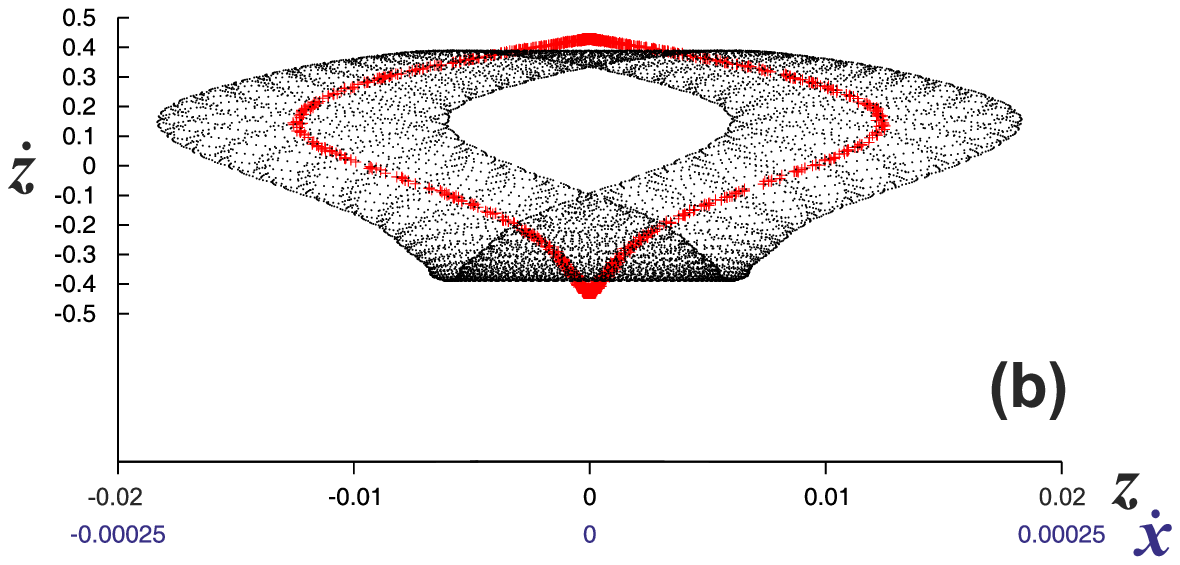}}\\
\end{tabular}
\caption{(a) The $(x,\dot x,z)$ and (b) the $(\dot x, z,\dot z)$ projections of the first 2580 consequents belonging to ``u''(red) and one of the outermost rotational tori around x1 (black). The point of view is given by the angles
$(\theta, \phi) = (90^{o},42^{o})$.}
\label{x1stick}
\end{center}
\end{figure}

The origin of the observed structure can be understood best in the ($z,\dot z$) projection (Fig.~\ref{x1v1v2}). We remind that we study the structure of phase space in the region U, which starts with the bifurcation of the family x1v2 at the $U\rightarrow S$ transition of x1 in the vertical 2:1 resonance region (Fig.~\ref{dsimple1}). For smaller energies in the resonance region, at $E_j=-5.028$, we have the $S\rightarrow U$ transition of x1 and the introduction in the system of the stable family x1v1 (and its symmetric family x1v1$^\prime$). Thus for $E_j=-4.66$, we have, apart from the two simple unstable p.o. x1v2 and x1v2$^\prime$, the two stable x1v1 and x1v1$^\prime$ and the planar stable x1. The presence of all these simple periodic orbits determines the orbital behavior of ``u'' in the following way:

In the central part of Fig.~\ref{x1v1v2}a we observe a red ``ellipse with
corners'', which forms within the first 2580 intersections and which is the
projection of the structure of Fig;~\ref{dsimple3} in the ($z,\dot z$)
plane. The initial conditions of x1v2 and x1v2$^\prime$ are located at the
upper and lower corner of the red elliptical curve respectively. The blue
ellipse inside the red one is the projection of a rotational torus
(Vrahatis et al. 1997) belonging to a quasiperiodic orbit   found by perturbing x1 in the $\dot z$ direction. During the first 2580 consequents the orbit ``u'' stays sticky to the outer 4D rotational tori of x1. 
The outermost rotational tori belonging to x1 are broader than the ribbon-like tori of the ``u'' orbit. In Fig.~\ref{x1stick} we give two characteristic projections  from the point of view $(\theta, \phi) = ( 90^{o},42^{o})$ in spherical  coordinates. In (a) we have the $(x,\dot x, z)$ and in (b) the $(\dot x, z, \dot z)$ projection. In both of them we overplot the  2580 consequents of ``u'' (red) and one of the outermost rotational tori around x1 (black).

By continuing the integration of the orbit ``u'' for more than 2580 intersections we see that the next 550 intersections form a \rotatebox[origin=c]{90}{$\Theta \:$}-like structure with two lobes on the sides of the red elliptical structure (Fig.~\ref{x1v1v2}a, indicated with a couple of white arrows). These consequents surround the rotational tori around the stable p.o. x1v1 and x1v1$^{\prime}$, which are very thin and appear in this projection like invariant curves (green and magenta).  The rotational tori of these two families are still in the phase space region close to the $U$ p.o. x1v2, since they have been bifurcated at a nearby $E_j$ from x1 (Fig.~\ref{dsimple1}).  The initial conditions of x1v1 and x1v1$^{\prime}$ are  in the middle of the rotational tori, marked with black symbols. 

Considering more than 3130 consequents we find that the next 1370 intersections stay trapped in a ring in the $(z,\dot z)$ projection, which in its turn surrounds all structures we have found in the phase space up to now. We point to this ring with two black arrows in Fig.~\ref{x1v1v2}a (right above and left below). 
Between the two lobes surrounding the rotational tori around x1v1 and x1v1$^{\prime}$ and the ring structure, we observe a chain of eight stability islands (colored black), which are the $(z,\dot z)$ projections of the corresponding rotational tori of a stable 8-periodic orbit.
Finally, after 4500 intersections, the consequents start diffusing into a larger volume of the phase space. Figure \ref{x1v1v2}a depicts the space of section after 5000 intersections. After 15000 intersections the consequents have occupied all available phase space (Fig.~\ref{x1v1v2}b). However, the denser regions clearly show us where the orbits have been trapped for longer times. 

We have to compare the structures described up to now with the 4D space of section we presented in Fig.~\ref{dsimple7}. The projections in Fig.~\ref{x1v1v2} are part of Fig.~\ref{dsimple7}. We observe them as we rotate the $(\dot x,z,\dot z)$ diagram in our computer screen. Since Fig.~\ref{dsimple7} consists of 3580 consequents, apart from the ribbon it includes also the consequents forming the  \rotatebox[origin=c]{90}{$\Theta \:$}-like structure around the rotational tori of x1v1 and x1v1$^{\prime}$ as well as part of the consequents that form the ``ring'' in Fig.~\ref{x1v1v2}. In Fig.~\ref{dsimple7} we emphasize on the distribution of the consequents in the 4th dimension. We observe that in Fig.~\ref{dsimple7} the color, representing the normalized $x$-coordinate values, is dominated by green shades. This means that although we do not have a smooth color variation as in Figs.~\ref{dsimple2},\ref{dsimple3} the consequents stay confined in a restricted volume of phase space. This happens indeed. The $x$ values for the first 3130 intersections are in the range $-0.008<x<0.08$, while within the next 450 intersections this range is almost doubled. Due to this expansion of the range of $x$ values, in Fig.~\ref{dsimple7} are introduced the orange-red and blue-violet consequents. The fact that in 3D projections we find structure and that the color variation is dominated by a  particular shade, indicates a weakly chaotic behavior of the orbit ``u''. This information is included in the 4D representation of the phase space as it is given in Fig.~\ref{dsimple7}. By combining Figs.~\ref{dsimple7} and \ref{x1v1v2} on a computer screen, we reveal not only the weakly chaotic character of the orbit, but also the regions where it sticks during the integration time. This shows the efficiency of the color-rotation method. 

\begin{figure}
  \begin{center}
\begin{tabular}{cc}
  \resizebox{80mm}{!}{\includegraphics{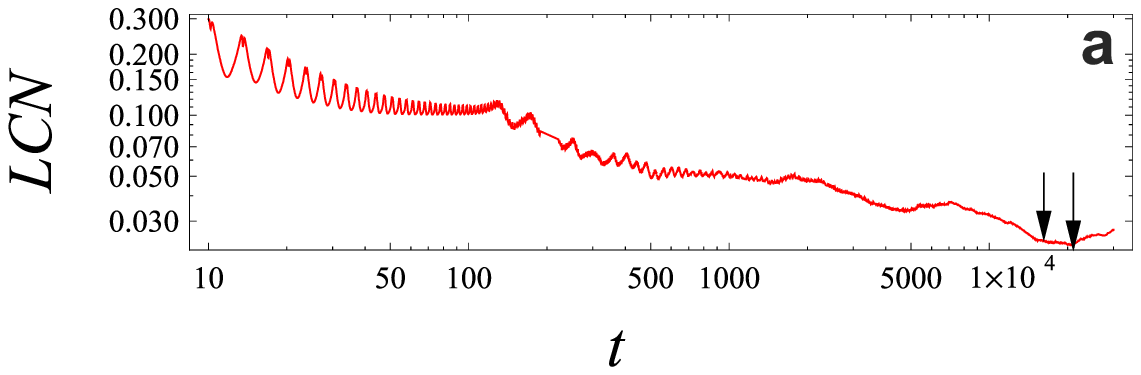}}\\
  \resizebox{82mm}{!}{\includegraphics{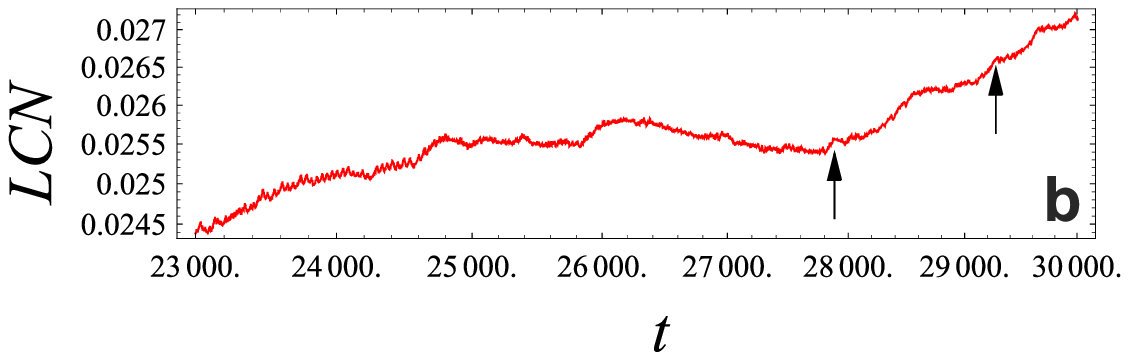}}\\
   \resizebox{80mm}{!}{\includegraphics{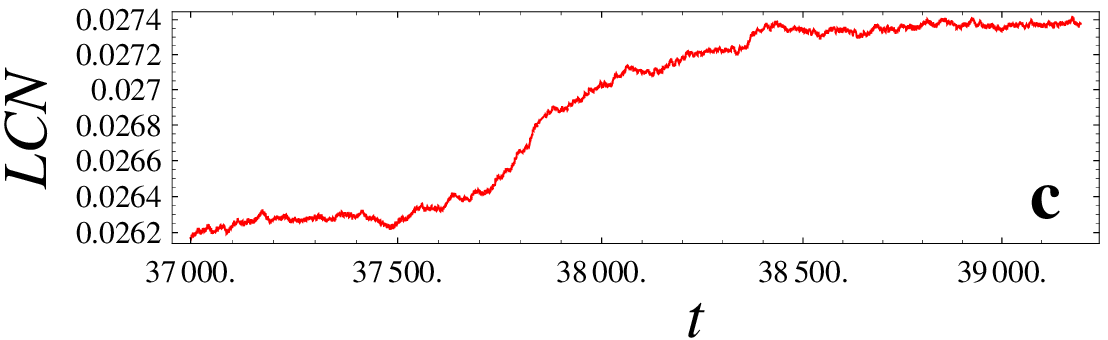}}
\end{tabular}
    \caption{The variation of $LCN(t)$ for the orbit ``u''. The ``ribbon''  4D morphology lasts until the point indicated with the left arrow in (a). The thin layer around the x1v1 and x1v1$^{\prime}$ rotational tori that corresponds to the lobes (Fig.~\ref{x1v1v2}a) is formed within the time interval between the two arrows in (a). The left arrow in (b) indicates a time at which we have an expansion of the consequents observed in the $(x, \dot x)$ projection, while to the right of the right arrow in (b) the consequents start visit all the available phase space. (c) Finally the curve levels off at a value $LCN(t) \approx 2.74 \times 10^{-2}$.}
    \label{lcn}
  \end{center}
\end{figure}

One can calculate for the orbit ``u'' the finite-time Lyapunov Characteristic Number, $LCN(t)$, which is defined by 
\begin{displaymath}
LCN(t)=\frac{1}{t}\ln\left|\frac{\xi(t)}{\xi(t_0)}\right|,
\end{displaymath}
where $\xi(t_0)$ and $\xi(t)$ are the distances between two points of two 
nearby orbits at times $t = 0$ and $t$ respectively (see e.g. Skokos 2010).
As we can see in Fig.~\ref{lcn} the changes we have found in the orbital behavior of ``u'' do correspond to characteristic changes of the slope of the $LCN(t)$ curve.
During the first 2580 intersections, indicated by the left arrow in Fig.~\ref{lcn}a the curve decreases towards zero. Then during the next 450 intersections, until the time pointed by the right arrow in Fig.~\ref{lcn}a when the consequents form the \rotatebox[origin=c]{90}{$\Theta \:$}-like structure with the lobes around the x1v1 and x1v1$^{\prime}$ tori, it continues to decrease but the curve has a much smaller slope. After that it starts increasing and its evolution after this point until $t=3\times 10^4$ can be followed in Fig.~\ref{lcn}b (note the different scales in the vertical axes). $LCN(t)$ increases, then levels off, and has a decreasing part until the left arrow in Fig.~\ref{lcn}b. which corresponds to about 4180 intersections. At this point we have an expansion of the consequents in phase space, which can be understood in the $(x,\dot x)$ projection (not shown in a figure). 
Beyond the left arrow in Fig.~\ref{lcn}b the $LCN(t)$ curve increases. 
Then, between Fig.~\ref{lcn}b and \ref{lcn}c the curve decreases and in Fig.~\ref{lcn}c it increases again.
Finally, for longer times of integration, when the orbit explores all the available phase space, the curve levels off at a value $LCN(t) \approx 2.74 \times 10^{-2}$ (Fig.~\ref{lcn}c)

The right arrow in Fig.~\ref{lcn}b corresponds to 4500 intersections and the  region to the right of this arrow  corresponds to the expansion of the consequents from the ring to the whole available volume of phase space as we observe in Fig.~\ref{x1v1v2}b. Although $LCN(t)$ gives globally the variation of the volumes of the phase-space filled by the consequents, by means of the color-rotation representation we can see in detail the specific regions where the consequents are trapped and the coordinate along which they diffuse to occupy larger volumes of the phase space. This information is essential in Galactic Dynamics for building self-consistent models using the orbital theory.

Summarizing the investigation of the orbital behavior of ``u'' in the neighborhood of a simple unstable periodic orbit we underline that it is characterized by two sticky periods, one associated with the rotational tori of x1 and a second sticky period associated with the rotational tori of x1v1 and x1v1$^{\prime}$. During the latter sticky period we observe the  \rotatebox[origin=c]{90}{$\Theta \:$} structure and the ring in the $(z, \dot z)$ projection. In the following section we examine how general is this behavior.

\subsection{The general orbital behavior in region ``U''.}

A dynamical behavior similar to that of the orbit ``u'' is
observed by perturbing the $\dot z_0$ initial condition of x1v2 up to $\Delta
\dot z \approx -10^{-2}$ (i.e. an absolutely largest $|\Delta \dot
z|=10^{-2}$). We find a similar behavior if we perturb $\dot x$ by at most
$\Delta \dot x \approx 2 \times 10^{-4}$.  Initial conditions that have the
same $x_0$, $\dot x_0$ and $z_0$ with x1v2 and $\dot z_0$ that deviates by
$\Delta \dot z < -10^{-2}$, give rotational tori around x1 (see Katsanikas \&
Patsis 2010).

Another class of orbits in the neighborhood of x1v2 is characterized by immediate stickiness around the x1v1 and x1v1$^\prime$ rotational tori, without building initially a ``ribbon''. We find them by keeping the same energy constant ($E_j=-4.66$) and   computing orbits by adding the perturbations given in Table 1 to the initial conditions of x1v2. In all  
these cases we do not observe  the well defined ribbon-like  surface that we have seen in  Figs. \ref{dsimple2} and \ref{dsimple3}. The orbits, starting close to the p.o. x1v2, depart directly from its immediate neighborhood and stick around the rotational tori of x1v1 and x1v1$^\prime$, before they expand and explore all available phase space. 
They form \rotatebox[origin=c]{90}{$\Theta \:$}-shaped structures surrounded by rings as in Fig.~\ref{dsimple7} without the central elliptical structure we observe in Fig.~\ref{x1v1v2}.
In Table 1 we give the ranges  
of the  perturbations of the initial conditions for which we have observed this orbital behavior.  For deviations from the p.o. in the $\dot z$ direction in the interval $4.4 \times 10^{-1}<\Delta \dot z < 4.7 \times 10^{-1}$ we reach the eight tori of the stability islands (Fig.~\ref{x1v1v2}).  By perturbing an initial condition we keep all others equal to the corresponding initial condition of the p.o.

\begin{table}
\begin{center}
\tbl{The range of deviations from the initial conditions of the periodic orbit x1v2
  for $E_j=-4.66$ that give orbits  whose consequents stay sticky to x1v1 and x1v1$^{\prime}$ and then diffuse occupying all the available phase space.}
{\begin{tabular}{|c|c|c|} \hline 
& positive direction & negative direction\\ \cline{1-3}
&&\\
$\Delta x$ & $\Delta x \leq 2\times 10^{-1}$ & $\Delta x \geq -2 \times 10^{-1}$\\ \cline{1-3} 
&&\\
$\Delta \dot x$ & $2\times 10^{-4}\leq \Delta \dot x \leq 1.2 $ & $\Delta \dot x \geq -2$\\ \cline{1-3}
&&\\
$\Delta z$ & $ \Delta z \leq 7\times 10^{-2}$ & $ \Delta z \geq -7\times 10^{-2}$ \\ \cline{1-3}
&&\\
$\Delta \dot z$ & $\Delta \dot z \leq 1.2 $ &-\\ \hline
\end{tabular}} 
\end{center}

\end{table}

For perturbations larger than those in Table 1 in the $x$, 
$\dot x$ or $\dot z$-direction  we do not find any stickiness but the consequents form clouds around the initial conditions of the p.o. x1v2. These clouds are not confined in a volume smaller than the total available phase space. They occupy always larger volumes as the integration time increases. We have to take into account that for larger $\Delta z$ perturbations than the listed in Table 1 we first fall on initial conditions of the rotational tori of x1v1 and x1v1$^{\prime}$ and then for even larger perturbations we find again the chaotic sea.
 
The second way we followed to explore the structure of phase space 
around x1v2, is by varying $E_j$, keeping the same perturbation in its initial 
conditions. We did so by adding a $\Delta x=10^{-4}$ perturbation in its $x_0$
initial conditions, for $E_j$'s in the region U. For $-4.7<E_j<-4.4$ the orbital 
dynamics of the orbits are as in the case for $E_j=-4.66$ that we described in 
section 4.1. However, for $E_j>-4.4$  a cloud of points surrounds 
the initial conditions of x1v2 in the 4D phase space.  This means that in all 3D projections we observe clouds surrounding x1v2 from the beginning of the calculation and also a mixing of colors.


\subsection{Asymptotic Curves}
Let us now try to understand the orbital behavior of the orbits we have studied in the previous section by investigating the structure of the asymptotic curves of x1v2 in the 4D spaces of section. The asymptotic curves are computed by mapping a line segment in the space of section on an associated dilating eigenvector, 
starting very close to a simple unstable periodic orbit x1v2 (see
e.g. Magnenat 1981).
The example we present is again  for $E_j=-4.66$. We have already 
studied the structure of the phase space in the neighborhood of this simple 
unstable periodic orbit in sections 4.1 and 4.2. The initial conditions of the p.o. 
have been given in section 4.1. 

We consider 
the unstable eigenvector\\ 
$(x_1,\dot x_1,z_1,\dot z_1)=
(-0.0011989,-0.00109617,-0.0459267,0.998943)$ 
and the corresponding  eigenvalue $\lambda=1.63536$. We compute the 
asymptotic curves of the unstable manifold and we study the morphology of its
asymptotic curves. The initial segment is composed 
of the initial conditions $(x_2,\dot x_2,z_2,\dot z_2)$  of the form:
 
\begin{equation}
\begin{split}
x_2=x_0+cx_1 \\
\dot x_2=\dot x_0+c\dot x_1 \\
z_2=z_0+cz_1 \\
\dot z_2=\dot z_0+c\dot z_1  
\end{split}
\end{equation}

First we compute the asymptotic curves for $c<0$ and we call this part of the 
unstable manifold, part A. The initial segment for  part A is 
composed of $10^{4}$ points (for $-10^{-3} \leq c \leq -10^{-7}$ with a step 
$10^{-7}$). For each point, 50 consequents were calculated. The basic 
criterion  for choosing the values of the perturbations and the number of the 
consequents is to stay  on the asymptotic curves with sufficient accuracy.
The monodromy matrix conserves its symplectic character (determinant $=1$) with
an accuracy at least of the order of $10^{-6}$. 

\begin{figure}[t]
\begin{center}
\begin{tabular}{cc}
\resizebox{90mm}{!}{\includegraphics{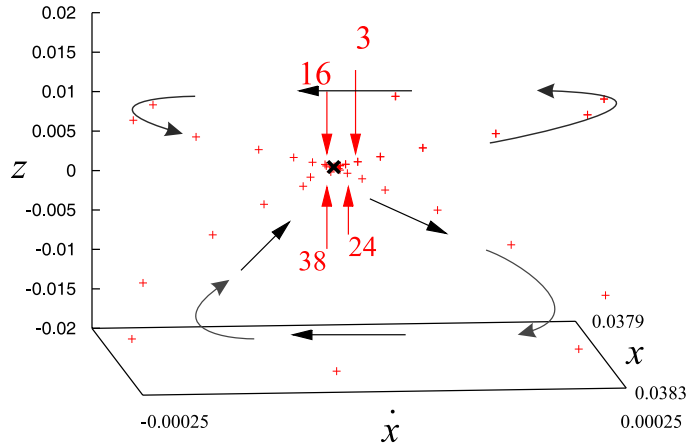}}\\
\end{tabular}
\caption{The first 50 consequents of  a single initial condition with $c=-10^{-3}$ in
  part A of the unstable manifold in the $(x,\dot x,z)$  projection. The  
  periodic orbit (p.o.) is indicated with a black symbol $\times$. The 
  red arrows point to the 3rd, 16th, 24th and the 38th consequent.  The 
  successive consequents (red crosses) follow the directions of the black 
  arrows. Our point  of view in spherical  coordinates is 
  $(\theta, \phi) = ( 78^{o},84^{o})$.}
\label{dman1}
\end{center}
\end{figure}

In Fig. \ref{dman1} we observe the first 50 consequents of part A of the
unstable manifold in the $(x,\dot x,z)$ subspace of the 4D  space of section.
The first 2 consequents are very close to the periodic orbit. Then, we discern 
the 3rd consequent (indicated with an arrow). The next 13 consequents follow 
the direction of the black arrows, starting to the right and upwards, and  they form a loop. After that, we return back to the neighborhood of the periodic orbit. 
Then, the 16th consequent as well as the following 7 remain again very close 
to the periodic orbit. From the 24th consequent and for the next 14 consequents  
we observe the formation of another loop, below the first one, following the 
directions that are indicated by the arrows. The final 12 consequents 
are again very close to the periodic orbit. The evolution in time of an initial segment of initial conditions leads to the formation of a
ribbon-like double loop structure, similar to the one in Fig. \ref{dsimple2}. 
 The same holds for the $(\dot x,z,\dot z)$ 3D projection of the manifold, which has a similar structure with the corresponding projection of the orbit ``u'' (Fig. \ref{dsimple3}). As in Figs.~\ref{dsimple2},\ref{dsimple3} color on the asymptotic curves varies smoothly, so the 4D morphology is as for the orbit ``u''.
 
\begin{figure}[t]
\begin{center}
\begin{tabular}{cc}
\resizebox{90mm}{!}{\includegraphics{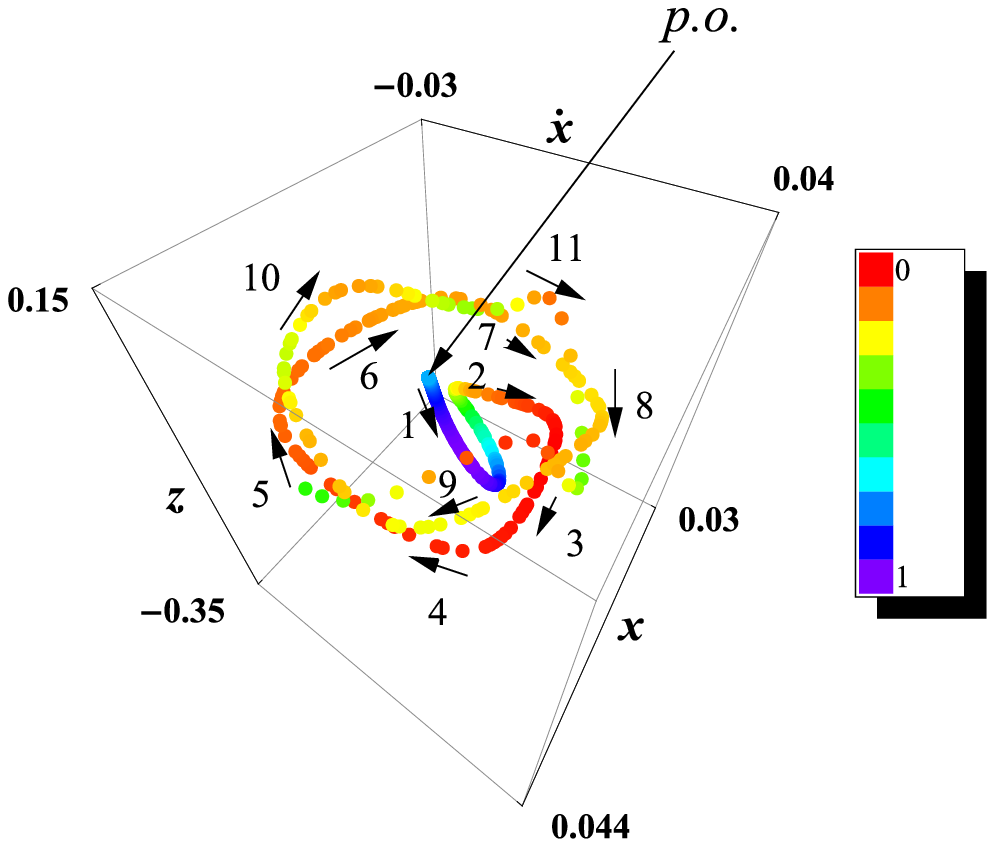}}\\
\end{tabular}
\caption{The 4D space of section  for the evolution of the first 60 points of the initial 
  segment of  part B in the case of the unstable manifold. The points are 
  depicted in the $(x,\dot x,z)$ subspace and are colored according their 
  $\dot z$ value. The successively numbered arrows indicate the direction of 
  evolution of the consequents along the manifold. The location of the  
  periodic orbit (p.o.) is indicated by an arrow. Our point  of view  in 
  spherical  coordinates is $(\theta, \phi) = (56^{o}, 22^{o})$.}
\label{dman6}
\end{center}
\end{figure}

For  $c>0$ in the initial segment of the initial conditions the asymptotic
curves evolve in the way we describe below and form the second part of the unstable manifold.
We call this second part, part B. The initial segment of initial conditions  for part B is 
composed by $2 \times 10^4$ points (for $c=10^{-6} \leq c \leq 2\times 
10^{-2}$ with a step $10^{-6}$). For every point we computed this time 20 
consequents. In Fig.~\ref{dman6} we depict the first 60 points of 
the initial segment in the $(x,\dot x,z)$ projection and color them according to their $\dot z$ values. The figure helps us understand that the asymptotic curves of part B of the unstable manifold are 4D curves, which wind around the point corresponding to the x1v2 p.o. in the space of section. Note that we have a smooth color variation along the curve  from light blue  to red, following the direction of the successive numbered arrows. 

\begin{figure*}[t]
\begin{center}
\begin{tabular}{cc}
\resizebox{95mm} {!}{\includegraphics{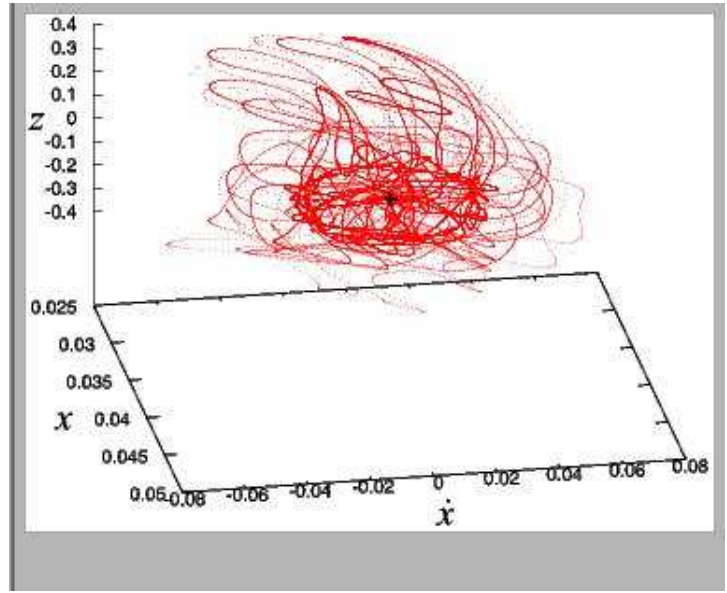}}\\
\end{tabular}
\caption{Part B of the unstable manifold in the 3D subspace $(x,\dot x,z)$ of the  space of section is given by following  the evolution of $2\times 10^4$ points in the initial segment for 20 consequents 
  (a total of $4\times 10^5$ consequents). The periodic orbit is indicated by 
  a black ``+''. Our point  of view in spherical  coordinates is in both panels
  $(\theta, \phi) = (56^{o},80^{o})$. }
\label{dman7a}
\end{center}
\end{figure*}

 Evolving the orbits for a larger number of points (consequents) from the initial segment (400 instead of 60 points) we observe that the asymptotic curves of part B oscillate rather far in the $(x, \dot x,z)$ projection (Fig. \ref{dman7a}).  For example in the $z$-direction these
oscillations extend between $-0.4<z<0.4$.  Considering the evolution of all
 $2 \times 10^4$ consequents of the initial segment the asymptotic curves fill 
roughly the 4D space $(x_1,x_2)\times(\dot x_1,\dot x_2) \times (z_1,z_2) \times(\dot z_1,\dot z_2)$
$= (0.025,0.05) \times (-0.08,0.08)\times(-0.4,0.4)\times(-1,1)$ (Fig. \ref{dman7a}). The figure is complicated and one can discern apart from the formation of loops at the upper left corner in Fig. \ref{dman7a}, a ring/disk structure that has at its center the initial conditions of the p.o.

As in the case of the orbits we have discussed in subsections 4.1 and 4.2, we also study the asymptotic curves by combining several 3D projections in which we color their points according to the fourth coordinate. Using the $(\dot x,z,\dot z)$ 3D projection and coloring the points according to their $x$ values the asymptotic curves are depicted in the following way.

\begin{figure}
\begin{center}
\begin{tabular}{cc}
\resizebox{65mm}{!}{\includegraphics{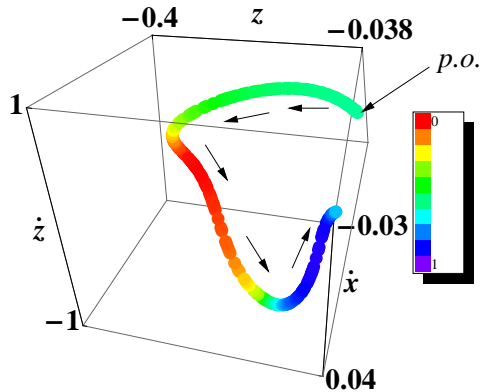}}
\end{tabular}
\caption{The first 20 points of the initial segment of part B of the unstable 
  manifold evolved for 20 intersections (a total of 400 consequents) in the 4D space of 
  section. The points are depicted in the $(\dot x,z,\dot z)$ subspace and are 
  colored according its $x$ value. The arrow labeled  with ``p.o.'' indicates 
  the location of the periodic orbit (p.o.). The consequents form a  loop 
  following the journey that is indicated with black 
  arrows. Our point of view in spherical  coordinates is 
  $(\theta, \phi) = (30^{o}, 45^{o})$.
}
\label{dman1011}
\end{center}
\end{figure}

In  Fig.~\ref{dman1011} we give the first 400 consequents totally (20 points on the initial segment evolved for 20 intersections). They
form an asymptotic curve that oscillates in the 3D 
$(\dot x,z,\dot z)$ projection. Along the asymptotic curve the color variation is smooth (from light blue to red to 
dark blue and then back to light blue), which means that the asymptotic curve is a 4D curve. Evolving further the asymptotic curve in time we observe that the oscillations become more complicated but the smooth color variation continues along the curve.
This clearly shows that the 
oscillations, which have been calculated by Magnenat (Magnenat 1982) in 2D projections, occur in the 4D space. 

\begin{figure*}[t]
\begin{center}
\begin{tabular}{cc}
\resizebox{90mm} {!}{\includegraphics{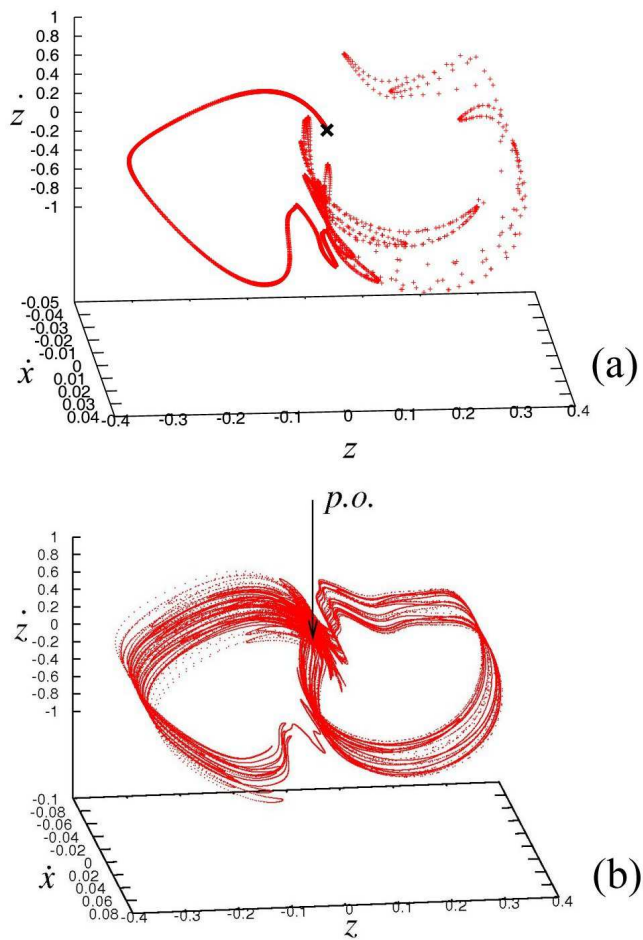}}\\
\end{tabular}
\caption{The asymptotic curves in the 3D subspace $(\dot x,z,\dot z)$.
  (a) the first 400 points of the initial segment of  part B of the unstable 
  manifold for 20 consequents (total 8000 consequents). The periodic orbit is 
  given with a black $\times$ symbol. After performing a loop to the left side 
  of the figure the asymptotic curves make several oscillations in the central 
  and the right part of the figure. The $\dot z$ values of the curve extend 
  roughly within a $\Delta \dot z=2$ range. Our point of view in spherical  
  coordinates  is $(\theta, \phi) = (68^{o}, 85^{o})$.
  (b) for $2\times 10^4$ points of the initial segment of  part B of the 
  unstable manifold for 20 consequents (a total of $4\times 10^5$ 
  consequents). The location of the periodic orbit (p.o.) is indicated with an 
  arrow. Our point of view in spherical  coordinates is 
  $(\theta, \phi) = (66^{o}, 84^{o})$.}
\label{dman12a}
\end{center}
\end{figure*}

In order to study the large scale evolution of part B of the unstable manifold we evolve next 400 points of the initial segment. The result is depicted in Fig.~\ref{dman12a}.
After one oscillation on the left side of Fig.~\ref{dman12a}a  we have several 
oscillations in the right part of the figure. The oscillations become longer 
than those of the previous case  (Fig.~\ref{dman1011}) and they extend to larger absolute values of $\dot z$ in the 
3D subspace $(\dot x,z,\dot z)$ (Fig. \ref{dman12a}a). By evolving 20000 points of the initial segment, the asymptotic curves form a \rotatebox[origin=c]{90}{$\Theta \:$}-shaped surface in the $(\dot x,z,\dot z)$ projection (Fig. \ref{dman12a}b) consisting of many parallel filaments. The filamentary character of this structure is due to the parallel oscillations of the asymptotic curves in the $(\dot x,z,\dot z)$ subspace.

\begin{figure}
\begin{center}
\begin{tabular}{cc}
\resizebox{90mm}{!}{\includegraphics{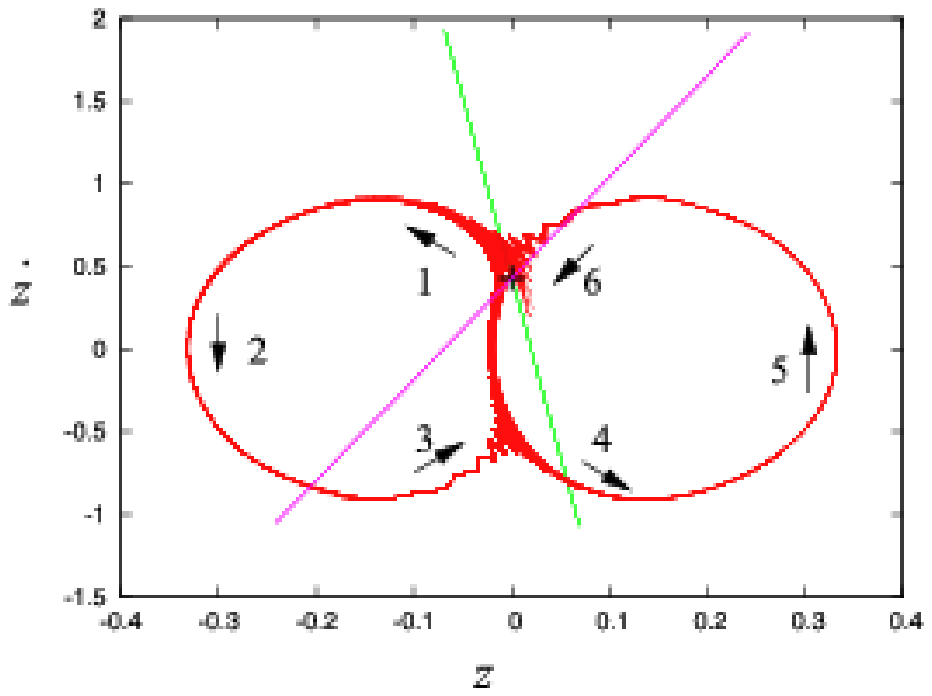}}\\
\end{tabular}
\caption{The 2D projection $(z,\dot z)$ of  part B of the unstable manifold 
  in the  space of section for 20000 points as in Fig. \ref{dman12a}b. The 
  arrows show the direction followed by the consequents. The stable and 
  unstable  eigendirections are depicted with magenta and a green lines 
  respectively. The periodic orbit x1v2 is indicated with a black  ``+'' symbol.}
\label{dman14}
\end{center}
\end{figure}

There is an obvious similarity of Fig.~\ref{dman12a} with Figs.~\ref{dsimple7} and \ref{x1v1v2}.
The projection of the asymptotic curves in five of the six 2D projections show clouds of points that seem to fill densely the available phase space. However, in the $(z,\dot z)$ projection we get a clear insight of how the curves evolve in time.  In this projection 
(Fig. \ref{dman14}) we see that the asymptotic curves of part B of the 
unstable manifold depart from the p.o. in the unstable direction 
(arrow 1) and  then they follow  the arcs as indicated by the arrows 
2,3,4,5 and 6. The asymptotic curves  oscillate  along a direction vertical to 
the stable eigendirection. 


The morphology we encounter for the case of the stable manifold is similar 
to that of the asymptotic curves of the unstable invariant manifold.
Below we discuss the morphology of the asymptotic curves for other values of 
the Jacobi constant in the region U of simple unstable x1v2 periodic orbits.

\subsubsection{Energy variation and asymptotic curves}
We describe now the invariant manifolds in the neighborhood of a simple 
unstable periodic orbit for values of $E_j$ $-4.7< E_j < -4.62$. They have
again two parts. 

\begin{figure}
\begin{center}
\begin{tabular}{cc}
\resizebox{100mm}{!}{\includegraphics{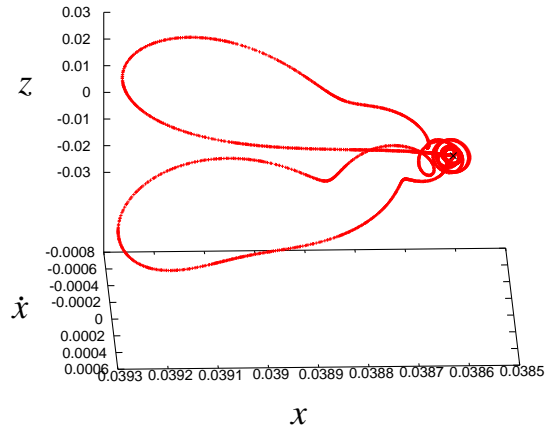}}\\
\end{tabular}
\caption{The $(x,\dot x, z)$ 3D projection of part A of the unstable 
manifold for $E_j = -4.63$. The asymptotic curves are computed for 10 
consequents and for 450 initial conditions (for  $ -4.5
\times 10^{-4} \leq c \leq -10^{-6}$ with a step $10^{-6}$ - a total of 4500 consequents). The 
location of the periodic orbit is indicated with a black ``$\times$'' symbol. Our 
point of view in spherical  coordinates  is  $(\theta, \phi) = (64^{o}, 
178^{o})$. }
\label{eman1}
\end{center}
\end{figure}

As the energy increases, for values larger than $E_j=-4.66$, the asymptotic 
curves of part A form  a ribbon, double loop, surface in the 3D 
subspace $(x,\dot x,z)$. As an example we give the case with  $E_j=-4.63$ (Fig.~\ref{eman1}). Although the lobes of the double loop are distorted, they form an acute angle in their $(x,z)$ projection like the corresponding lobes of the orbit ``u'' (lower middle panel in Fig.~\ref{dsimple2}). We observe also that the consequents wind around the 
periodic orbit (as the consequents in Fig.~\ref{dman6}) when they approach the 
periodic orbit. The manifold has a similar structure like the one we have encountered in the central region of ``u'' in the same 3D projection.
Also, in these higher energies, the asymptotic curves of  
part A form  again a circular  ribbon-like surface in the 3D subspace 
$(\dot x,z,\dot z)$ similar to the orbit in Fig.~\ref{dsimple3}. The increase of energy causes oscillations of the 
asymptotic curves of part A, in the 3D subspace $(\dot x,z, \dot z)$.

In the interval $-4.61\leq E_j \leq -3.47$ the invariant manifolds   
do not have two different forms. Along both directions the asymptotic curves have the
same morphology. The asymptotic curves  resemble  the asymptotic 
curves of  part B of the unstable invariant manifolds for $E_j=-4.66$ (the 
case that we described in the section 4.3).   

\subsection{Stickiness}
From the investigation of the spaces of section and the numerical calculation of the asymptotic curves we presented in the two previous subsections it becomes evident that the phenomenon of stickiness not only is present but characterizes the dynamical behavior of many orbits that start in the neighborhood of the simple unstable orbit x1v2. As we 
know,  stickiness in chaos refers to chaotic orbits that stay in a particular region of the phase space, before escaping to large distances. In 2D autonomous Hamiltonian systems stickiness appears near the borders of stability islands, as well as near the  asymptotic curves of 
the unstable manifolds for a long time (Contopoulos \& Harsoula 2008, 
Contopoulos \& Harsoula 2010). In our 
paper we  explore this phenomenon in the 4D space of section of our 3D system and we try to explain it in combination with the structure of the associated manifolds. It is the first 
time that stickiness  in chaos is studied in detail in the neighborhood of simple
unstable periodic orbits of a 3D autonomous Hamiltonian system. In order to study the phenomenon of stickiness in our system we consider first the representative example  of   a periodic orbit in the
region U of the x1v2 family for $E_j=-4.66$.

\begin{figure}[t]
\begin{center}
\begin{tabular}{cc}
\resizebox{100mm}{!}{\includegraphics{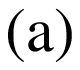}}\\
\resizebox{100mm}{!}{\includegraphics{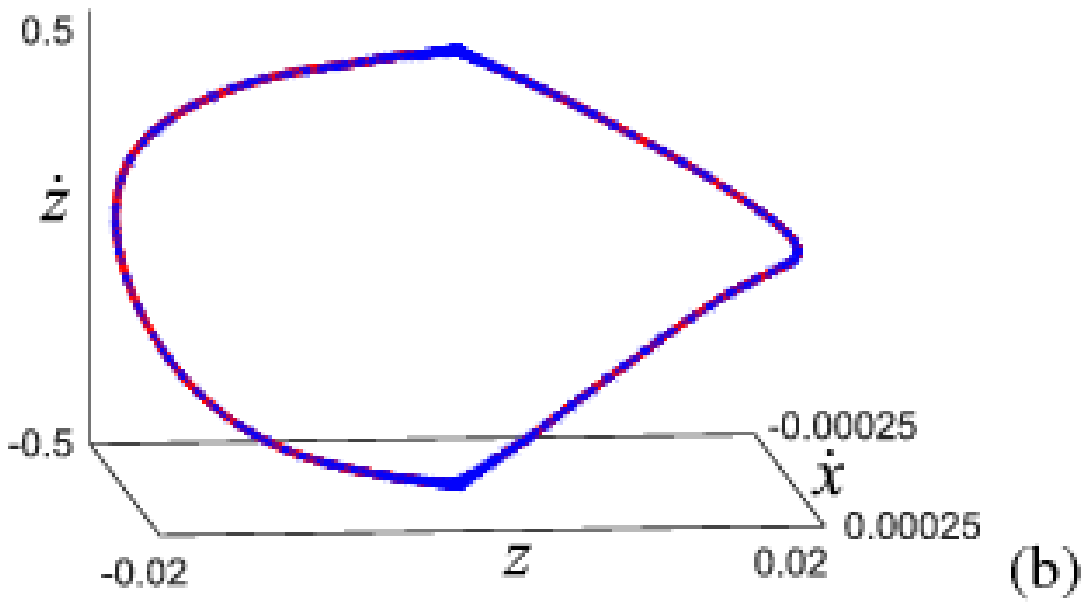}}\\
\end{tabular}
\caption{The first 2580 consequents of the orbit ``u'' (with blue color)
  sticky to the asymptotic curves  of the part A of the unstable 
  manifold (with red color) that surround the x1 rotational tori. (a) in the 3D subspace $(x,\dot x,z)$.  
  Our point of   view in spherical  coordinates  is $(\theta, \phi) = (78^{o}, 
84^{o})$. (b) in the 3D subspace $(\dot x,z,\dot z)$ with point of view in 
  spherical  coordinates $(\theta, \phi) = (74^{o}, 82^{o})$ }
\label{dstic1}
\end{center}
\end{figure}

If we add a negative 
perturbation in the $\dot z$ direction (as we mentioned in section 4.2), or add  a small positive
perturbation in the $\dot x$ direction (for $10^{-6} \leq \Delta \dot x < 
2 \times 10^{-4}$) to the initial conditions of the periodic orbit x1v2,  
the orbits we find have a dynamical behavior similar to the one of the orbit ``u'' 
(section 3.1). By comparing the orbital behavior of all orbits having a  morphology like ``u'', with the form of the unstable manifolds we conclude that the initial conditions of ``u'' are very close to the unstable eigendirection of the manifold. Both the consequents of ``u'' and the asymptotic curves wind around the rotational tori associated with x1. The consequents of ``u'' follow the asymptotic curves and stick in the region dictated by the manifolds.
For example in  Figs. \ref{dstic1}a and \ref{dstic1}b the first 2580 consequents of the orbit ``u'' (depicted by blue color) stick 
close to the asymptotic curves of  part A of the  unstable manifold (drawn 
with red color). This number of consequents correspond to a time of 124 
periods of rotation of our system, a very large time interval in galactic dynamics (of the order of several Hubble times).

\begin{figure}[t]
\begin{center}
\begin{tabular}{cc}
\resizebox{80mm}{!}{\includegraphics{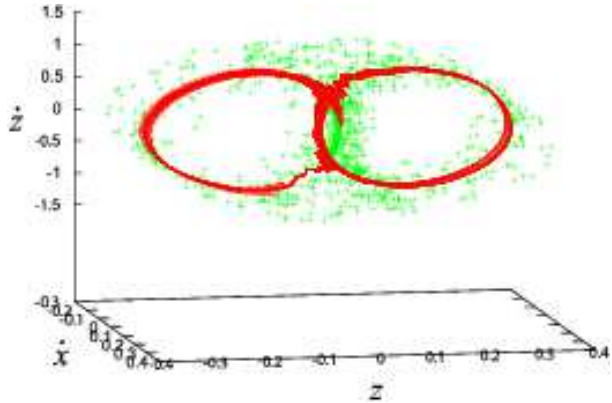}} 
\end{tabular}
\caption{The first 1000 consequents  of the orbit ``u'', given with green ``+'' 
  symbols  after their  departure  from   part A of the unstable 
  manifold, as observed  in the 3D projection $(\dot x,z,\dot z)$. These 
  consequents stick close to the asymptotic curves of  part B of the unstable 
  manifold (drawn with red color). Our point of view in spherical  
  coordinates  is $(\theta, \phi) = (78^{o}, 79^{o})$. }
\label{dstic3}
\end{center}
\end{figure}

For a larger number of consequents, the points depart from part A of the  
unstable manifold. The relation between the consequents of ``u'' and the unstable manifold is best understood again in the 
3D projection $(\dot x,z,\dot z)$. As we can see in this projection in Fig.~\ref{dstic3}, the first 1000 
consequents, after the departure of the consequents from  part A of the 
unstable manifold, remain close to the asymptotic curves of part B of the 
unstable manifold. This is the mechanism that creates in phase space the lobes
of the \rotatebox[origin=c]{90}{$\Theta \:$}  and the ring structure we
observe as denser regions in the $(z, \dot z)$ projection in
Fig.~\ref{x1v1v2}. These denser regions are created due to the phenomenon of
stickiness around the rotational tori of x1v1 and x1v1$^{\prime}$. The lobes
of the \rotatebox[origin=c]{90}{$\Theta \:$} structure, indicated with white
arrows in Fig.~\ref{x1v1v2}, are formed by consequents that stay almost on the
(red) asymptotic curves of Fig.~\ref{dstic3}. However, even after the
departure of the consequents from the lobes they stay close to the x1v1 and
x1v1$^{\prime}$ rotational tori, because they are trapped in the immediate
neighborhood of the eight islands of stability we observe as small black line
segments in Fig.~\ref{x1v1v2}.  In this projection the figure resembles what
is found in 2D systems close to the last KAM curve (Contopoulos 2002). However, we should have in mind that in the 4D space of section of our 3D Hamiltonian system consequents can be found on both sides of a torus, moving through the additional dimensions. 

In section 4.2 we said that for the values of perturbation listed in Table 1,
the orbits go directly around the rotational tori of x1v1 and x1v1$^{\prime}$. All these orbits have also a sticky behavior at the borders of the x1v1 and x1v1$^{\prime}$ 4D stability islands, which we can always see best in the 3D projection $(\dot x,z,\dot z)$ and the 2D projection $(z,\dot z)$.  In all other projections the sticky regions are discernible as denser regions of the phase space embedded in clouds. The $(\dot x,z,\dot z)$ projection has the advantage that in a certain range of viewing angles only a few consequents from the cloud are projected in the regions, which look ``empty'' (actually they are occupied by the rotational tori around the stable orbits).

Our study of the dynamical behavior in the neighborhood of the simple unstable orbits, shows that the characteristic double loop structures that have been already found by Magnenat (1982) and Patsis and Zachilas (1994) etc., as well as the \rotatebox[origin=c]{90}{$\Theta \:$}-shaped structures and rings we observe in Fig.~\ref{x1v1v2}, are associated with stickiness at the borders of rotational tori of stable families that exist in the system. The consequents follow the manifolds that wind around the tori.  Thus it would be more accurate to associate these structure of the phase space mainly with the $U\rightarrow S$ transitions, as they are not expected to exist away from such a transition point or after a $DU\rightarrow U$ transition.

\section{Double Instability}

Double instability in rotating galactic potentials usually appears for large
energies. In the x1 orbital tree (Skokos et al 2002) $DU$ appears in higher order bifurcations, when a stable 3D bifurcation of x1 becomes first simple unstable and then double unstable, or when a simple unstable bifurcation of x1 becomes double unstable. Frequently we find $DU$ p.o. at energies for which nearby orbits escape. Thus the selection of the appropriate case for investigating the phase space structure is crucial. Here we consider first the dynamics in the neighborhood of  
orbits of the x1v2 family in the region DU (Fig.~\ref{dsimple1}). However, in
order to examine the dependence of the dynamical behavior on the energy of the
orbit, we study also the phase space close double unstable 
z-axis orbits. These are p.o.  along the rotation axis z in rotating triaxial 
potentials (Heisler et al. 1982). 

\subsection{x1v2}

The periodic  orbits of the x1v2 family are DU for large energies 
(Fig. \ref{dsimple1}). At first we  study the case for $E_j=-3.38$. The  
periodic orbit has initial conditions $(x_0,\dot x_0,z_0,\dot z_0)=
(0.088998826,0,0,2.3614474)$. By  adding even a tiny  perturbation 
(e.g. $\Delta x=10^{-6},10^{-5}$ or$10^{-4}$) to its initial conditions, along 
any direction, we  find clouds  of points in all 3D projections of the 
4D space of section 
These are 4D clouds of points since we find scattered points in all 3D 
projections and mixing of colors when we apply the method of color and 
rotation. Their consequents occupy a volume in  the  4D space of  section, 
lying roughly in the region $(x_1,x_2) \times(\dot x_1,\dot x_2) \times (z_1,z_2)\times(\dot z_1,\dot z_2) 
= (-4,4)\times(-3,3)\times (-2.5,2.5)\times (-3,3)$.

Then we explore the structure of the phase space  in the neighborhood of 
the  double unstable periodic orbits of x1v2 in the region DU from 
$-3.41 < E_j < -2.3$, for the same fixed value of the  perturbation,  $\Delta 
x=10^{-4}$. We find  again 4D clouds of points as in the case of the double 
unstable periodic orbit at $E_j=-3.38$.   

\begin{figure}[t]
\begin{center}
\begin{tabular}{cc}
\resizebox{100mm}{!}{\includegraphics{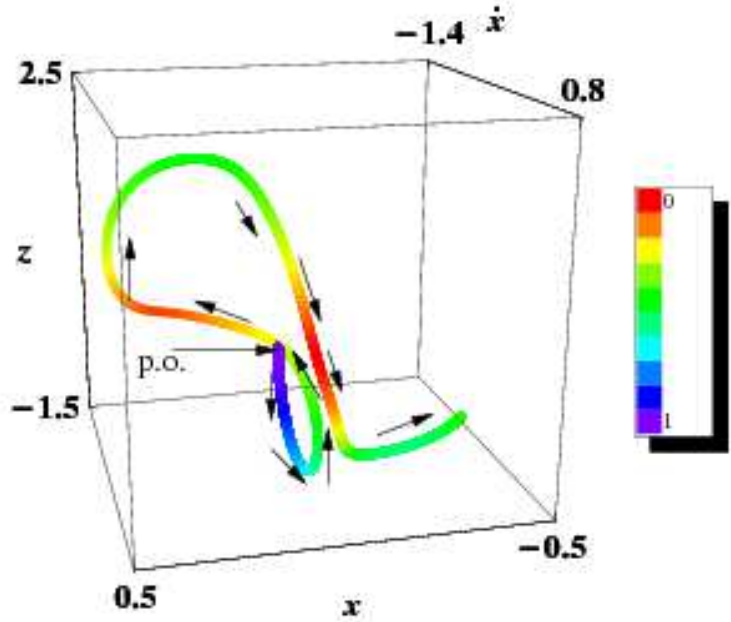}}\\
\end{tabular}
\caption{The  asymptotic curves  of the 
unstable invariant manifold in case ``I'', for 3 consequents of 2000 
initial conditions of the initial segment with $c_2=0$ (a total of 6000 
consequents), in the neighborhood of the double unstable 
periodic orbit for  $E_j=-3.38$ in the 4D space of section. The consequents 
are depicted in the $(x,\dot x,z)$ subspace and are colored according to their 
$\dot z$ value. The location of the p.o. is  indicated with a long arrow.
The consequents depart from the p.o. in the direction pointed by arrows. The point  of view in 
spherical  coordinates  is  $(\theta, \phi) = (45^{o}, 180^{o})$.}
\label{dunst2}
\end{center}
\end{figure}

We  study  the morphology of the asymptotic curves of the 
unstable manifold in the neighborhood of the double unstable periodic orbits
in the region DU  and  we present this morphology  in the neighborhood of the 
double unstable periodic orbit for  $E_j=-3.38$ (Fig. \ref{dunst2}).  The 
computation of the asymptotic curves  is similar to that of the case of the 
simple unstable periodic orbits (section 4.3). The only  difference is that we 
have now two unstable eigenvectors. These  are $(x_1,\dot x_1,z_1,\dot z_1)= 
(0.0216618,0.163481,-0.737238,0.655198)$  and $(x_2,\dot x_2,z_2,\dot z_2)
= (-0.831661,0.523481,0.0173103,0.184412)$. The corresponding eigenvalues are
$\lambda_1= 69.08801$ and $\lambda_2= 1.74981$. The initial conditions 
$(x_3,\dot x_3,z_3,\dot z_3)$, that we  use for the
computation of the asymptotic curves, are given by: 
 
\begin{equation}
\begin{split} 
x_3=x_0+c_1 x_1+c_2 x_2\\
\dot x_3=\dot x_0+c_1 \dot x_1+c_2 \dot x_2\\
z_3=z_0+c_1 z_1+c_2 z_2\\
\dot z_3=\dot z_0+c_1 \dot z_1+c_2 \dot z_2\\
\end{split}
\end{equation}

We computed the asymptotic curves, in two cases which we named ``I'' and
``II'' respectively. In the first case, we kept  $c_2$  constant and equal to 
zero and  we used $c_1$ as a parameter. In the second case, we kept  $c_1=0$  
and  we varied $c_2$. In  case ``I'' we have taken values $c_2=0$ and  
$10^{-7} \leq c_1 \leq 2\times 10^{-4}$ with a step $10^{-7}$ 
(2000 initial conditions). In the second case ``II''  we  took values $c_1=0$ and 
we varied $c_2$ in the interval  $-1.35\times 10^{-1} \leq c_2 \leq -10^{-5}$ 
with a step $10^{-5}$ (13500 initial conditions). In both cases ``I'' and 
``II'' we computed only 3 consequents for every initial 
condition\footnote{ During the 
computation of asymptotic curves, we connect gaps that appear along them by considering more
initial conditions in these gap regions.}. Here we present the morphology of the asymptotic curves in 
the 3D $(x,\dot x,z)$  projection.  We found in our numerical experiments, that  the morphology  of the 
asymptotic  curves in  all 3D projections is similar. As it was expected the
consequents along the asymptotic curves  in  case ``I'' (with the large 
eigenvalue $\lambda_1$) move away faster from the periodic orbit  than  the
consequents along the asymptotic curves in case ``II'' (with the small 
eigenvalue $\lambda_2$).

The consequents  of the unstable
invariant manifold in case ``I'' depart from  the periodic orbit by 
following  the black arrows of Fig.~\ref{dunst2}. They 
form first a  loop in the 3D projections of 
the 4D space of section below the periodic orbit. Then the consequents 
approach  again the periodic orbit but with a different color (yellow) than
the one of the initial part which is violet. Afterwards  the asymptotic curve 
 performs a larger loop before leaving the box of the frame.
In addition we observe a smooth color
distribution  from violet (close to the p.o.) to blue, green, yellow and red 
and then follows a color oscillation from red to green and back again. These oscillations correspond to oscillations of $\dot z$ in  the upper part of the color spectrum (right part of Fig.~\ref{dunst2}). This smooth color variation along the curve indicates that these asymptotic curves are 4-dimensional curves.  

\begin{figure}[t]
\begin{center}
\begin{tabular}{cc}
\resizebox{100mm}{!}{\includegraphics{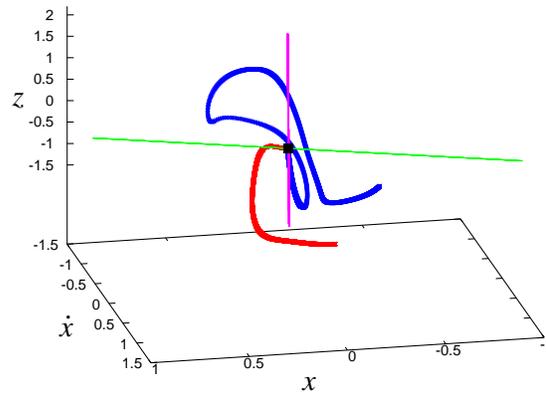}}\\
\end{tabular}
\caption{The 3D subspace $(x,\dot x,z)$ of the  asymptotic curves  of the 
unstable invariant manifold for case ``I'' (blue color) and case ``II'' 
(red color), in the neighborhood of the double unstable periodic orbit for  
$E_j=-3.38$ in the 4D  space of section. The two unstable eigendirections that
correspond cases ``I'' and ``II'' are depicted with a magenta and a green
line respectively. The location of the periodic orbit 
is given with a black square symbol. The point of view in spherical  
coordinates  is  $(\theta, \phi) = (63^{o}, 168^{o})$.}
\label{dunst2a}
\end{center}
\end{figure}

In both cases ``I'' and ``II'' the asymptotic curves deviate soon from the eigendirections.
In  Fig.~\ref{dunst2a} we compare the 
asymptotic curves  ``I'' (blue color) and ``II'' 
(red color) that start along the unstable eigendirections that correspond to the
large $(\lambda_1)$ and small $(\lambda_2)$ eigenvalues respectively.
The morphologies of these curves are different. The orbits we start integrating close to these 
curves  remain sticky to them only for a short time (0.4 periods of rotation 
of our system) before moving away from them. Soon they  fill in a chaotic way 
a large region of the phase space.  We have found similar morphologies for all 
cases of stable and unstable manifolds we have examined in the DU region.

This is a dynamical behavior typical for $DU$ periodic orbits of the x1 tree. However, it is not necessarily typical for the dynamics in the neighborhood of $DU$ periodic orbits in general. We investigated the orbital behavior close to double unstable p.o. in several cases in our system. Below we describe the dynamics close to z-axis $DU$ p.o.'s, which differs in many aspects from what we have found in the neighborhood of the $DU$ x1v2 p.o.'s we presented up to now.

\subsection{z-axis}
As already mentioned, the orbits of the z-axis family lie entirely on the rotational, z axis, of
the system. Thus, we consider now the $z=0$ surface as surface of section with
  $\dot z >0$. Earlier studies have shown that this family becomes double
  unstable in slowly rotating triaxal systems (Martinet \& de Zeeuw 1988,
  Patsis \& Zachilas 1990). We
  investigated the stability of this family in the potential (2) rotating very slowly with $\Omega_b = 5\times 10^{-6}$. For this value of $\Omega_b$ the z-axis family has large regions of double instability (DU). In this case, as we vary the energy, the z-axis family for large negative values of $E_j$ is  stable and for $E_j=-6.7$ it
becomes simple unstable. At this transition bifurcates the stable
family of p.o. of stable anomalous orbits, abbreviated to sao 
(Heisler et al 1982), and its symmetric family  with respect the yz-plane. Then the z-axis
family, for $E_j=-5.836$,  becomes double unstable and it remains DU for larger values of $E_j$.

In the $(x,\dot x,y,\dot y)$ 4D space of section the initial conditions of the z-axis orbits are $(0,0,0,0)$. For $E_j=-4.9$ we present the case of the orbit in the neighborhood of the z-axis p.o., which deviates from it in the $\dot x$ direction by $\Delta \dot x=0.62$. We call this orbit ``duz1''.



\begin{figure}
  \begin{center}
\begin{tabular}{cc}
  \resizebox{80mm}{!}{\includegraphics{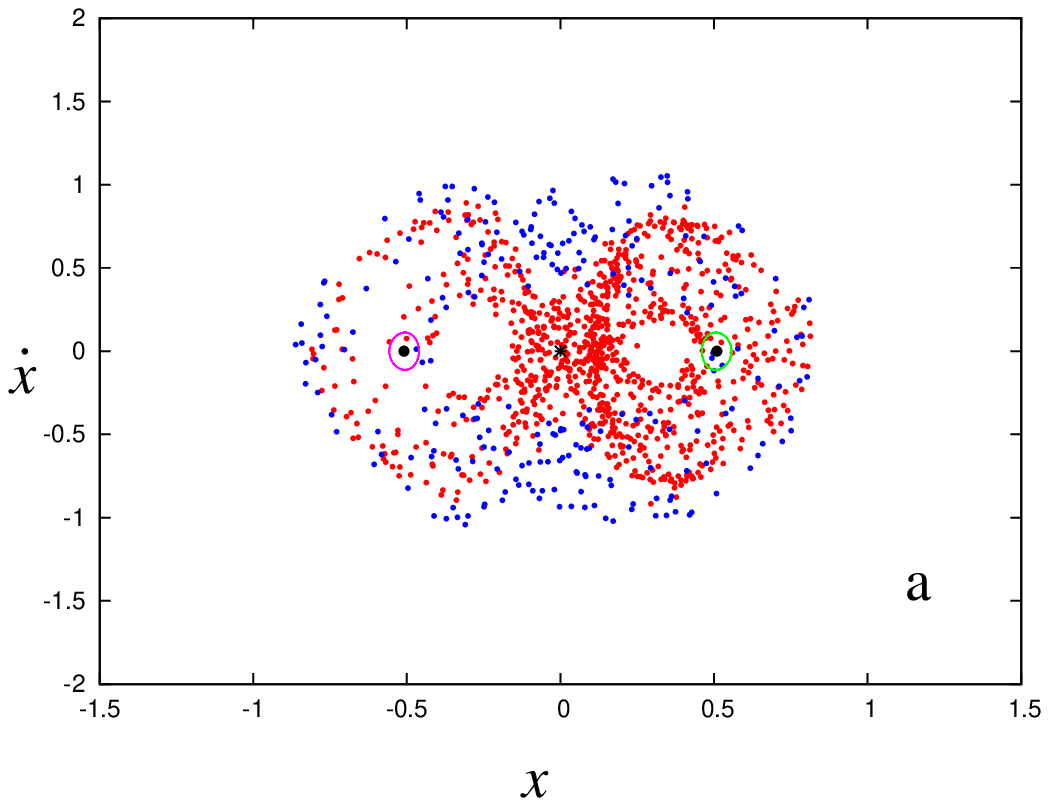}}\\
  \resizebox{80mm}{!}{\includegraphics{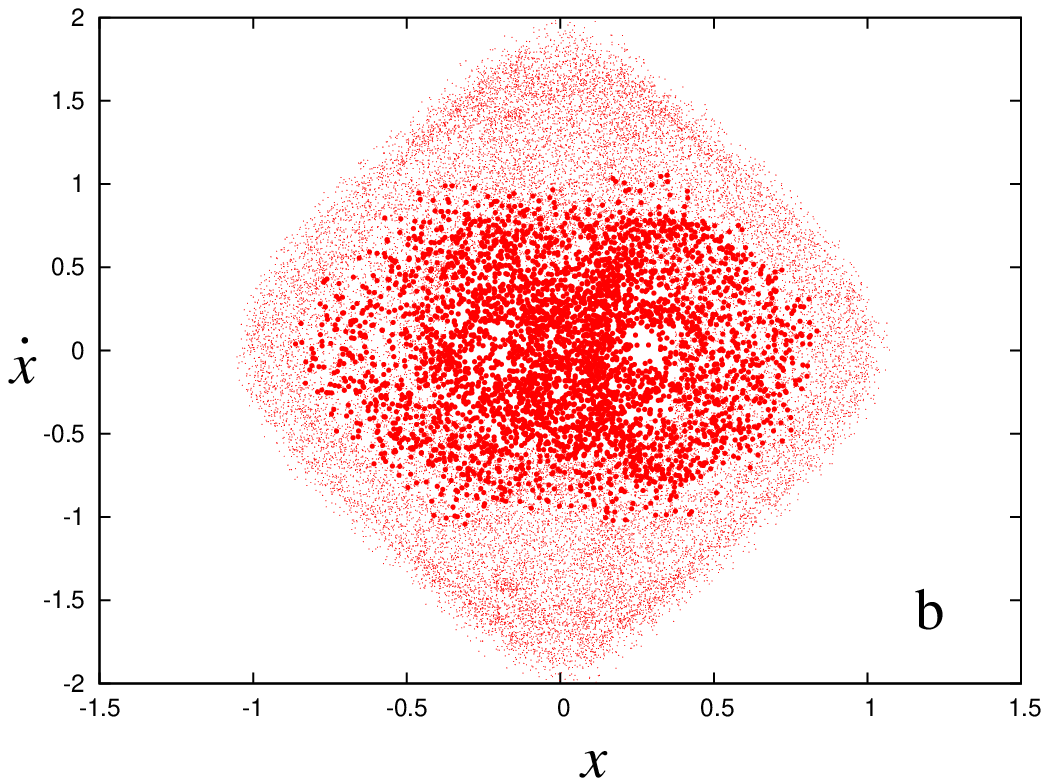}}\\
\end{tabular}
   \caption{(a) The 2D $(x,\dot x)$  projection of the orbit we study in the neighborhood of the $DU$ z-axis orbit at $E_j=-4.9$. We consider here 4200 consequents. The first 250 are plotted with  blue color. 
The next 3950 form the red ``$8$''-shaped structure. Then they start to diffuse in phase space. Black symbols indicate the location of the periodic orbits in this projection. (b) The first 4200 consequents are depicted now with heavy red points, while the next 26000 (light red dots) diffuse and occupy all available phase space.}
    \label{zdu1}
  \end{center}
\end{figure}

In the 2D $(x,\dot x)$ projection we observe the first 250 consequents being
distributed around the p.o. (blue points in Fig.~\ref{zdu1}a). These points do
not show clearly any  particular structure. The next 3950 consequents however, form an ``$8$''-figure. We plot with red color 950 of them in Fig.~\ref{zdu1}a. In the same figure 
the periodic orbits existing in this energy are given with black symbols. In
the center of the  ``$8$''-figure structure the $\times$ symbol indicates the
$DU$ z-axis orbit, while in the central regions of the two lobes of this
structure the black dots (on the left and the right)
indicate the location of sao and its symmetric p.o., which are surrounded by
projections of rotational tori around them. If we continue to integrate the
orbit, we find that after more than  4200 intersections the consequents diffuse  and occupy all available phase space. Then the $(x,\dot x)$ projection is as we see in Fig.~\ref{zdu1}b.

On one hand we have the formation of a structure discernible in projections of the phase space. On the other hand, the consequents that form the  ``$8$''-figure structure of Fig.~\ref{zdu1}a are not on a smooth 4D surface.
The application of the color-rotation method shows that we can hardly observe
a smooth color variation along any direction. These are indications of a  
chaotic behavior.

\begin{figure}
  \begin{center}
    \resizebox{100mm}{!}{\includegraphics{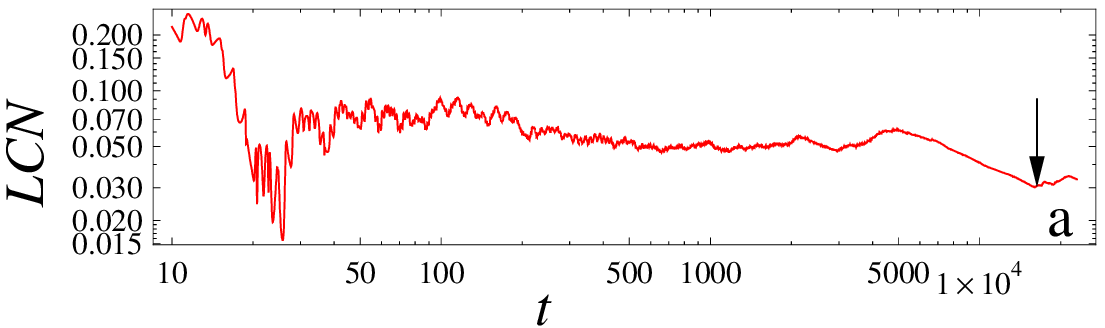}}
    \resizebox{100mm}{!}{\includegraphics{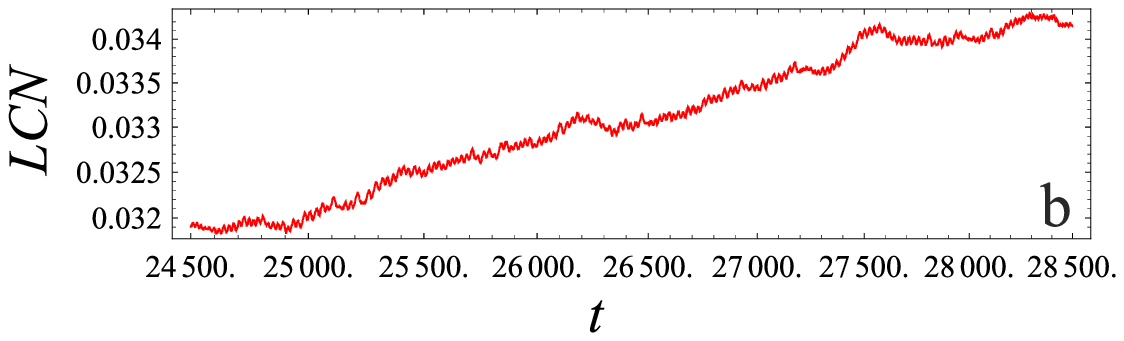}}
    \caption{The time evolution of the finite time $LCN$ of the orbit in Fig.~\ref{zdu1}. The arrow in (a) indicates the time at which the consequents depart from the ``$8$''-shaped structure that they formed during the first 4200 intersections. After that point $LCN(t)$ increases as we can see in (b).}
    \label{zlia}
  \end{center}
\end{figure}

The finite time $LCN$ index points also to the chaotic character of the orbit.
For most of the time when the consequents in the $(x,\dot x)$   (or the $(x,\dot x,y)$)
projection stay on the ``$8$''-shaped figure, the index varies around $5\times
10^{-2}$ and finally decreases until the time pointed by an arrow in
Fig.~\ref{zlia}a. This time corresponds to 4200 consequents. In
Fig.~\ref{zlia}b we see how $LCN(t)$ evolves as the consequents start filling
all available phase space. It increases continuously up to $t=28500$
and beyond.

The range of perturbations for which we find the  ``$8$''-figure is registered in Table~2. Outside the range given for the $x$ coordinate  we find orbits sticky to rotational tori around the sao and its symmetric orbit.  For absolutely larger  
perturbations in the $\dot x$, $y$ or $\dot y$-direction we encounter three types of orbits, namely (a) orbits represented by  clouds of points, (b) orbits sticky to tori around the stable periodic orbit x1v1 and (c) orbits sticky to chains of rotational tori  belonging to p.o. of higher multiplicity. This orbital behavior is related with the unstable manifold of the z-axis orbit as follows. 

\begin{table}
\begin{center}
\tbl{The range of perturbations  that give orbits whose consequents are
  distributed as in  Fig.~\ref{zdu1}.}
{\begin{tabular}{|c|c|} \hline
 $-0.2 \leq \Delta x \leq 0.2 $& $-1.5 \leq
  \Delta \dot x \leq 1.5 $\\ \cline{1-2}
$-0.7 \leq \Delta y \leq 0.7$& $-1 \leq \Delta \dot y \leq 1$\\
\hline
\end{tabular}}
\end{center}
\end{table}

The computation of the manifolds  is similar to that of the case of the double unstable periodic orbits
of x1v2. The unstable eigenvectors   are $(x_1,\dot x_1,y_1,\dot y_1)=
(-0.20427,0.000957614,0.978912,-0.00197504)$  and
$(x_2,\dot x_2,y_2,\dot y_2)=
(-0.000880093,0.436429,0.00422376,-0.899728)$.
The corresponding eigenvalues are $\lambda_1= 5.93858$ and
$\lambda_2= 5.90405$. The initial conditions  $(x_3,\dot x_3,y_3,\dot y_3)$,
which we  use for the computation of the asymptotic curves, are given by:
\begin{equation}
\begin{split}
x_3=x_0+c_1 x_1+c_2 x_2\\
\dot x_3=\dot x_0+c_1 \dot x_1+c_2 \dot x_2\\
y_3=y_0+c_1 y_1+c_2 y_2\\
\dot y_3=\dot y_0+c_1 \dot y_1+c_2 \dot y_2\\
\end{split}
\end{equation}

\begin{figure}[t]
\begin{center}
\begin{tabular}{cc}
\resizebox{150mm}{!}{\includegraphics{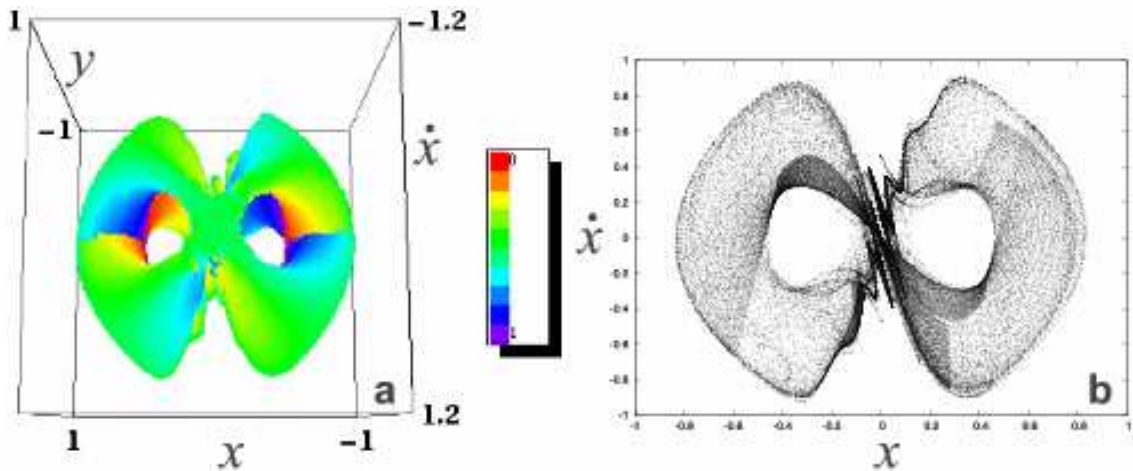}}
\end{tabular}
\caption{(a) The 4D representation of the unstable asymptotic surface, using the 3D projection $(x,\dot x,y)$ for the spatial representation of the consequents and $\dot y$ for the normalized color values. The point of view in spherical coordinates 
is  $(\theta, \phi) = (20^{o}, 190^{o})$. (b) The $(x,\dot x)$ projection showing the oscillations of the asymptotic surface close to the z-axis $DU$ p.o. located at (0,0).}
\label{zman0}
\end{center}
\end{figure}

We calculate the manifold by using for $c_1$ the values
 $-10^{-3} \leq c_1 \leq  10^{-3}$ with a step $10^{-5}$ and for every value of
$c_1$  we consider for $c_2$  the values  $-10^{-3} \leq c_2 \leq  10^{-3}$
again with a step $10^{-5}$. For every initial condition we compute 10 
consequents. Then we have totally $4\times 10^{5}$ consequents. This way we 
computed the unstable asymptotic surface we present in Fig.~\ref{zman0}.  We
note that now we have an asymptotic surface and not an asymptotic curve. This
is expected since we have two eigenvalues outside the unit circle (see
e.g. Arnold 1988, p. 287). The asymptotic surface is easier to be visualized than in the case of the orbit in the neighborhood of the double unstable x1v2 orbit, because now  the two associated eigenvalues are very close. 
We observe that the  unstable
asymptotic surface we calculated has a morphology of an  ``$8$''-figure
in the 3D projection $(x,\dot x,y)$ of the surface of section. For the 4th dimension we color the points according to the normalized $\dot z$ values. We have a complicated but smooth
color variation on the surface of the  ``$8$''-shaped structure (Fig.~\ref{zman0}a). In Fig.~\ref{zman0}b we give the $(x,\dot x)$ projection of the same figure that shows clearly the central oscillations of the surface close to the p.o. located at the center of this ``$8$''-figure structure.

A similar  morphology is found for the stable asymptotic surface. 

\begin{figure}[t]
\begin{center}
\begin{tabular}{cc}
\resizebox{80mm}{!}{\includegraphics{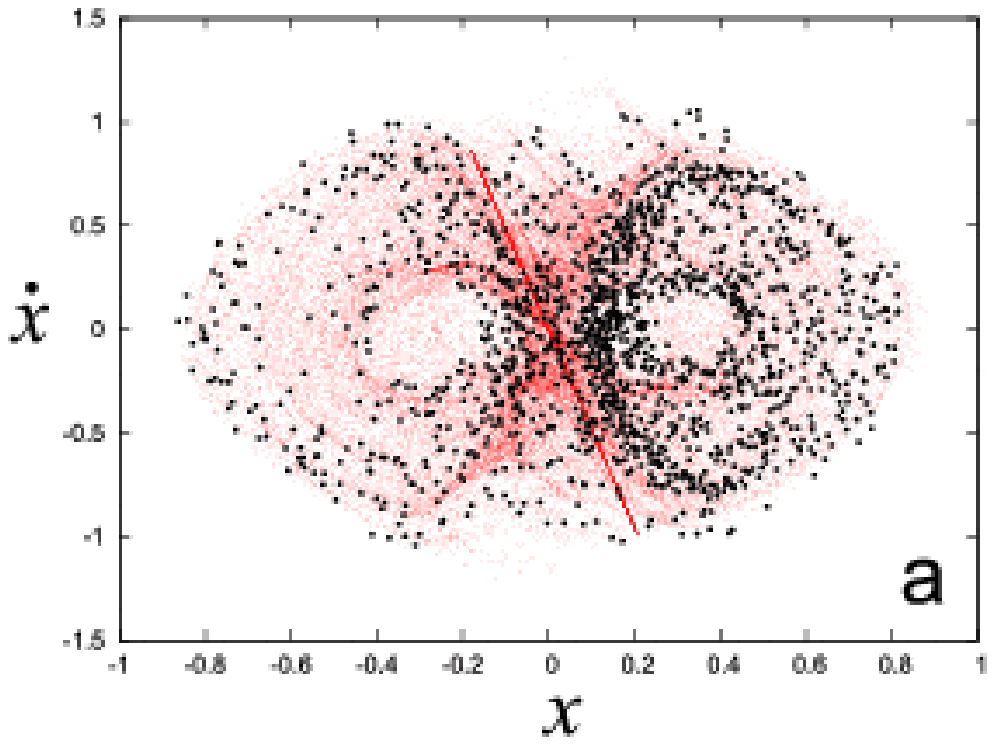}}
\resizebox{80mm}{!}{\includegraphics{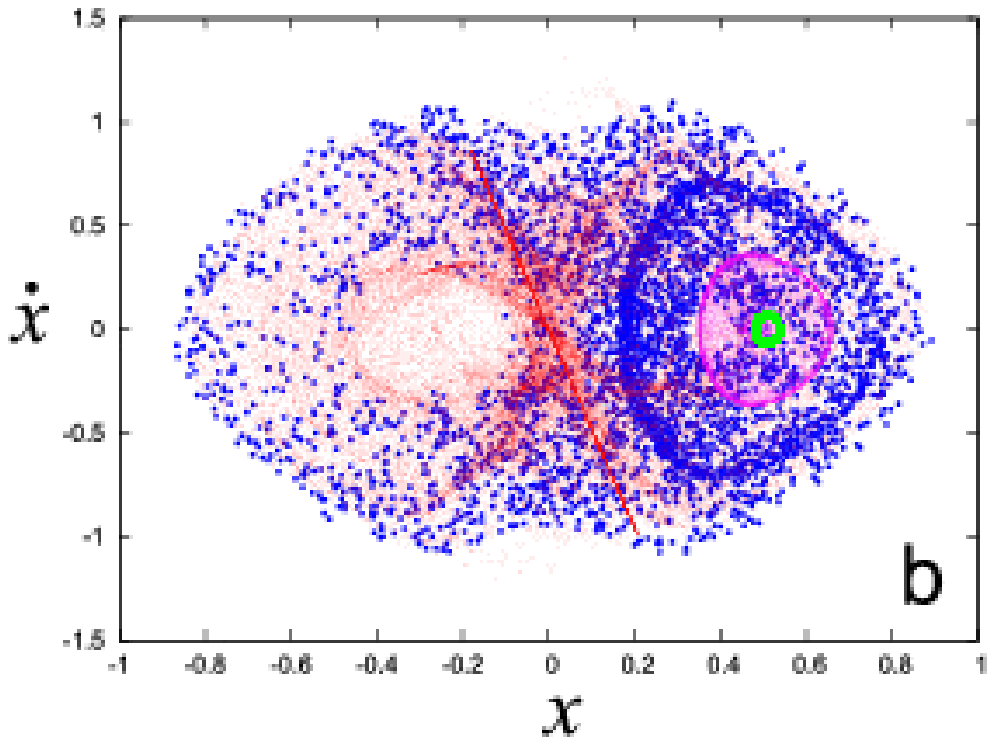}}
\end{tabular}
\caption{(a) The $(x,\dot x)$ 2D projection of the unstable asymptotic surface
  (red color) and the consequents of orbit ``duz1'' that remains for long times sticky to it. (b) The same projection of the unstable asymptotic surface, together with two orbits (blue and magenta consequents) that we find by perturbing orbits of the sao rotational tori (projected to a  small green elliptical curve near the center of the right part of the manifold).}
\label{zman1}
\end{center}
\end{figure}

The structure we present in Fig.~\ref{zman0} is part of the unstable manifold. If we consider a larger number of consequents for $-1\leqq c1,c2 \leqq 1$ and a step $10^{-3}$ in (5),  the asymptotic surface expands considerably also in the $y$-direction and its $(x,\dot x)$ projection is as the red points show in Fig.~\ref{zman1}a. We realize that \textit{all} consequents  of ``duz1''
(black color) stick on the unstable asymptotic surface.



We have to note also that inside the lobes of the ``$8$''-shaped manifold structure, besides the sao rotational tori (green elliptical curve in Fig.~\ref{zman1}b) we find also sticky orbits, initially forming toroidal objects in this region and then diffusing to a larger volume, remaining however sticky to the surface of the manifold. The consequents of two such orbits are given with blue and magenta colors in the right lobe of the manifold in Fig.~\ref{zman1}b. The  existence of the asymptotic surface decelerates the diffusion of these orbits to larger phase space regions for long times.



\section{Diffusion in phase space}

The study of the orbits in the neighborhood of simple and double unstable 
periodic orbits gives us an information of practical interest namely  their 
rate of diffusion and the possible trapping of their consequents in particular 
regions of the available phase space. This would allow us to separate the 
orbits that support physical structures from those which do not. For this 
reason we measure the diffusion  of the orbits we study and we compare the 
results among themselves.  

From the orbits close to x1v2 p.o. we have consider the following orbital 
behaviors:
\begin{itemize}
 \item The first type of orbits we found close to x1v2 periodic orbits is represented by double loop ribbons or by closed ribbons in the 3D projections 
of the 4D space of section. An example is the  orbit ``u'' presented in 
section 4.1. (Figs.~\ref{dsimple2},\ref{dsimple3}). The orbits of this orbital type 
continue beyond the initial ribbons, by forming additional dense regions in phase space after long 
integration times. These dense regions are  discernible in some projections as ``$8$''-like structures or rings
(Fig.~\ref{x1v1v2}). 

 \item The second type of 
orbital behavior is represented by clouds of points in the 4D surface of
section. As an example, we consider
the orbit with $\Delta x=10^{-4}$ from the initial conditions of the simple unstable x1v2 for $E_j=-4$ and we call it ``u1''.
\item The 
third type of orbital behavior is   represented again by clouds of points in the 4D 
surface of section, but this time in the neighborhood of a double unstable periodic orbit. 
As a typical example for the discussion to follow, we consider the orbit deviating from x1v2 by $\Delta x=10^{-4}$ at $E_j=-3.38$. We call this orbit ``du1''.
\end{itemize}
In addition we include an orbit in the neighborhood of a z-axis $DU$ p.o., i.e.
\begin{itemize}
 \item The orbit that builds an ``$8$''-shaped figure before diffusing and occupying all available volume of the phase space (Fig.~\ref{zdu1}), which we call ``duz1''.
\end{itemize}

In order to estimate the diffusion velocity for the above types of
orbital behavior we define a diffusion time and a diffusion velocity as follows:
In general the volume of phase space occupied by the consequents of a chaotic orbit increases with time until they occupy the maximum available volume restricted by the surface of the zero velocity.
The time  that the orbits need to occupy the 
maximum volume $V_{max}$, we define as ``diffusion time'' $T_d$. Then the  
``diffusion speed'' is the ratio \[u_d = \frac{V_{max}}{T_d}.\]
The  quantities $V_{max}$ and $T_d$ for the orbits in the neighborhood of a periodic 
orbit can be easily computed in two steps:

\begin{enumerate}
\item As time increases we compute the mean distance from the periodic orbit $s$ up to this time $t$ of all
consequents and then we compute a mean volume $\overline{V}(t)$. In other words we compute a ``mean volume'' of the hypersphere centered on the periodic orbit with radius the mean distance $s$, i.e the quantity:
$\overline{V}(t)=\frac{1}{2} \pi^{2} s^{4}$.
\item After time  $T_d$ the consequents occupy the maximum volume
  $V_{max}$. This is represented by the maximum  peak in a $\overline{V}(t)$ diagram and this allows us to
  compute the 
diffusion speed.
\end{enumerate}
  
\begin{figure}
  \begin{center}
\begin{tabular}{cc}
  \resizebox{80mm}{!}{\includegraphics{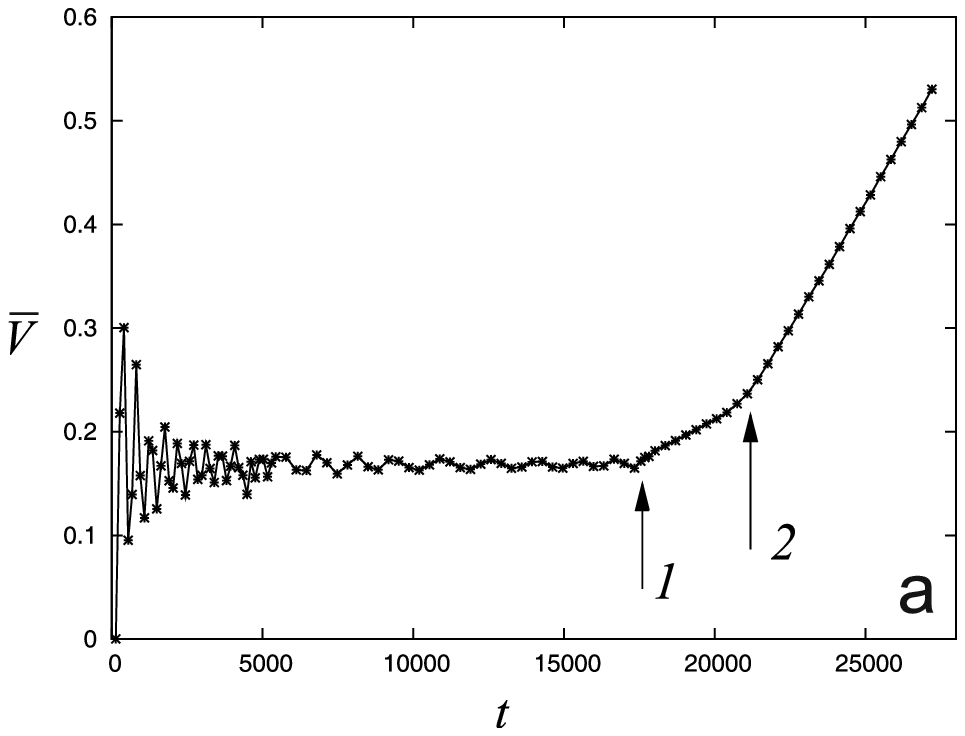}}\\
  \resizebox{80mm}{!}{\includegraphics{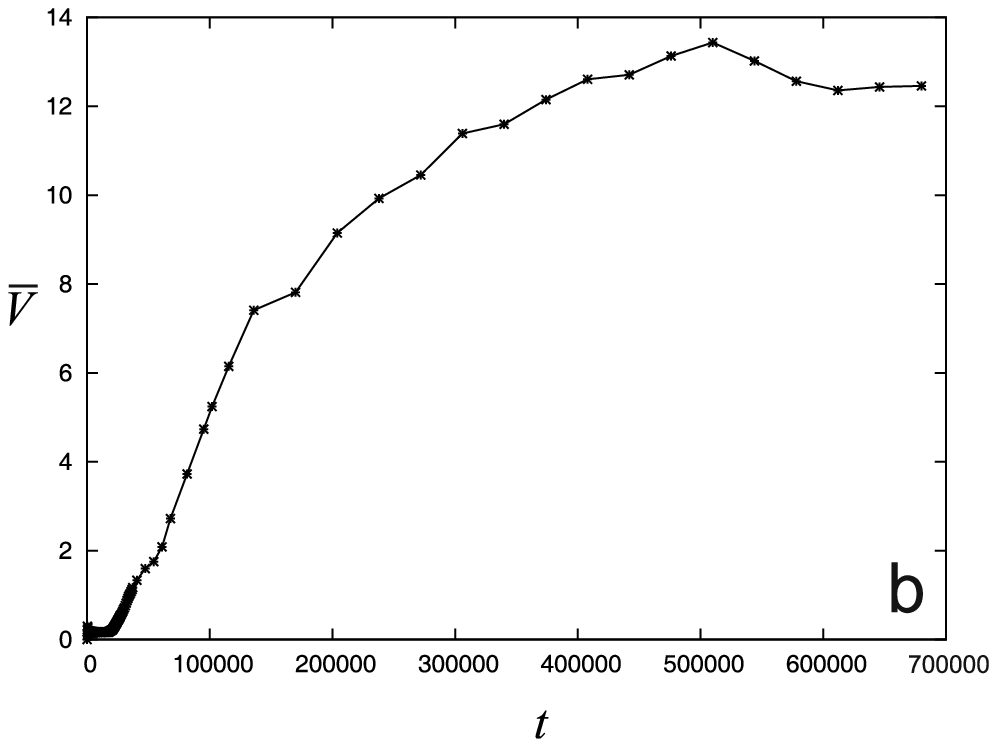}}\\
\end{tabular}
    \caption{The mean volume $\overline{V}(t)$ that is occupied by the consequents of orbit
      ``u'' in the neighborhood of the simple unstable periodic orbit for 
     $E_j=-4.66$. (a) $t \leq 28000$, corresponding to 4200 consequents. The arrows indicate the times corresponding to 2580 and 3130 intersections. (b) The evolution of $\overline{V}(t)$  up to  $t = 7 \times 10^5$.}
    \label{diff1}
  \end{center}
\end{figure}

In Fig.~\ref{diff1}a  we plot the volume occupied by the
consequents of the orbit ``u'' (Figs.~\ref{dsimple2},\ref{dsimple3}) versus time $t$ for an interval  $t \leq
28000$, corresponding to 4200 intersections. We observe that for  $t \lesssim 17300$ in our system units (indicated by the left arrow in Fig.~\ref{diff1}a corresponding to 2580 intersections) $\overline{V}(t)$ remains close to 0.165 after some initial oscillations. 
 This time interval
corresponds to the phase that the consequents stick to the asymptotic curves of 
part A of the unstable manifold forming the double loop ribbon (section 4.1).
Then we observe a second interval between the two arrows in Fig.~\ref{diff1}a, where $\overline{V}(t)$ increases but the slope of the curve is smaller than the slope beyond the right arrow.  The second arrow corresponds to time $t=21800$ or 3130 intersections.  The dynamical behavior between the two arrows 
corresponds to the time during which the consequents form  the lobes of the \rotatebox[origin=c]{90}{$\Theta \:$}-structure and the ``ring'' in the $(z, \dot z)$ projection (Fig.~\ref{x1v1v2}) sticking on the asymptotic curves of part B of the 
unstable manifold (section 4.4). 
Beyond the right arrow $\overline{V}(t)$ increases continuously with a characteristically steeper slope than before. The consequents leave all structures observed in the neighborhood of x1v2 (section 4.1).  Figure \ref{diff1}b is a continuation of Fig.~\ref{diff1}a  and shows that the peak of $\overline{V}(t)$ corresponds to
$V_{max}=13.433$ for $T_d=510000$. The  speed of diffusion is  
$u_d=\frac{V_{max}}{T_d} =2.63 \times 10^{-5}$.
After the peak of $\overline{V}(t)$ in Fig.~\ref{diff1}b the consequents arrive at distances  
from the periodic orbit that are smaller or equal to the maximum distance. This means that $\overline{V}(t)$ decreases a little and finally, it levels off to a constant value.  This 
happens for all chaotic orbits. 

\begin{figure}
  \begin{center}
    \resizebox{100mm}{!}{\includegraphics{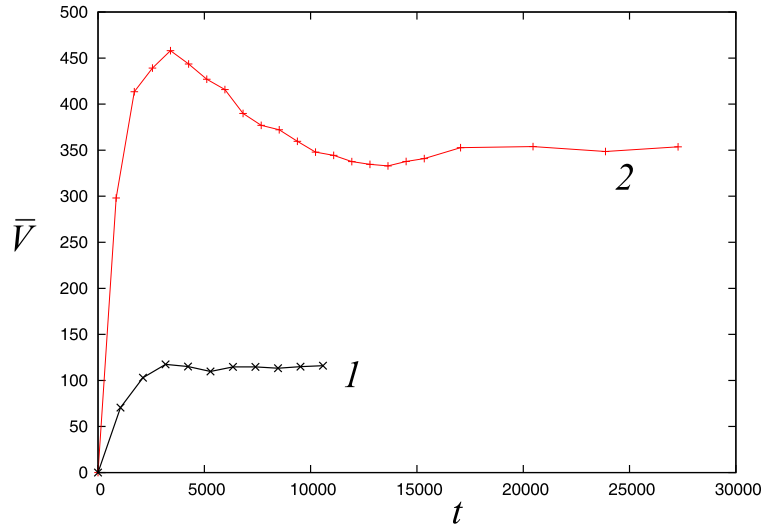}}
    \caption{The evolution of $\overline{V}(t)$ for the orbit
      ``u1'' in the neighborhood of the simple unstable periodic orbit x1v2
      for $E_j=-4$ (1) and for the orbit ``du1'' (2), when x1v2 became double 
      unstable for $E_j=-3.38$.}
    \label{diff2}
  \end{center}
\end{figure}

The evolution of $\overline{V}(t)$ and the quantity $T_d$ change in the case of the orbit ``u1''.
In Fig.~\ref{diff2} we see that for the curve labeled ``1'', corresponding to the orbit ``u1'', we have $V_{max}=117.397$ and $T_d=3180$. This means that the speed of diffusion is now $u_d=0.0369$, i.e. it
is 1403 times larger than the velocity of diffusion of orbit ``u''. 
The upper, red curve, labeled ``2'', describes the evolution of $\overline{V}(t)$ for the orbit ``du1'' in the neighborhood of the $DU$ p.o. x1v2 at $E_j=-3.38$. We observe that the curve has a maximum  
$V_{max}=458$ and $T_d=3400$. In this case the speed of diffusion is  
$u_d=0.135$ that is 3.65 times and 5133 times larger than the diffusion speeds of  the orbits ``u1'' and ``u'' respectively. For ``u1'' and ``du1'' $\overline{V}(t)$ increases rather fast to the peak of the diagram,  indicating that we do not have any important stickiness in these cases.
 
\begin{figure}
  \begin{center}
\begin{tabular}{cc}
  \resizebox{80mm}{!}{\includegraphics{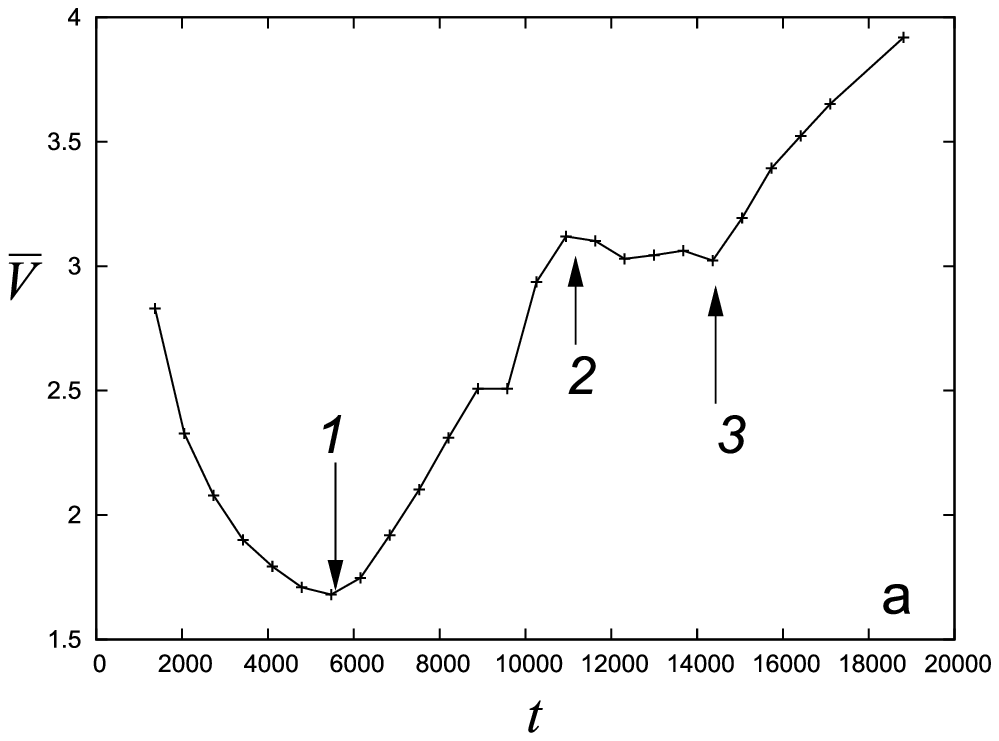}}\\
  \resizebox{80mm}{!}{\includegraphics{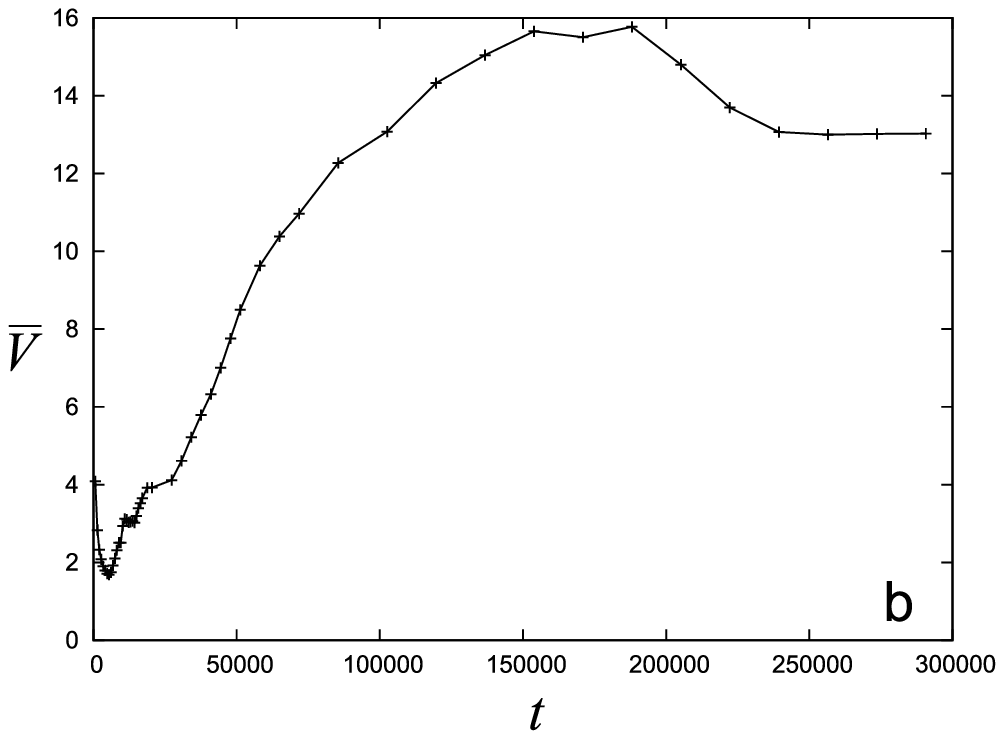}}\\
\end{tabular}
   \caption{The evolution of $\overline{V}(t)$ for orbit
      ``duz1''  in the neighborhood of the double unstable periodic orbit of the
      z-axis family for $E_j=-4.9$. (a) For $t \leq 2 \times 10^4$ and
      (b) for $t \leq 3\times 10^5 $.}
    \label{zdiff}
  \end{center}
\end{figure}

Finally for the case we have found stickiness in chaos close to a $DU$ z-axis p.o., i.e. for the orbit ``duz1'',  the evolution of the quantity $\overline{V}(t)$ is given in Fig.~\ref{zdiff}.
Initially $\overline{V}(t)$ decreases to a value about 1.68 during the first 1600 consequents. This period ends at the time indicated by arrow ``1''. This happens because the
majority of the points occupy the right lobe of the ``$8$'' structure, as in the
$(x,\dot x)$ 2D projection (Fig.~\ref{zdu1}). This implies that the majority of the  points are
included in a volume  smaller than  the volume occupied by the consequents forming the whole
``$8$''-shaped structure. Thus, the mean volume decreases. 
Then the next consequents complete the ``$8$'' structure by populating both lobes with similar number of consequents, so $\overline{V}(t)$ increases until $\overline{V}(t)= 3.11969$, which corresponds to 3200 intersections (arrow numbered ``2''). 
Then we observe 
a small plateau until the point  indicated by the arrow labeled  ``3''
($\overline{V}(t)=3.02259$). The time at this point corresponds to 4200 consequents. 
This is the number of
consequents that  stay on the eight-figure surface before they depart from it and occupy a larger volume
of the phase space. For larger times $\overline{V}(t)$ increases. In Fig.~\ref{zdiff}b we have a
peak at $V_{max}=15.77$ for $T_d=1.88 \times 10^5$. The
diffusion speed is $u_d=8.385\times 10^{-5}$.

It is evident by comparing Figs.~\ref{diff1}, \ref{diff2} and \ref{zdiff}, that the diffusion of chaotic orbits from the neighborhood of simple or double unstable p.o. does not depend on the kind of instability of the periodic orbit. The phenomenon of stickiness that decelerates the diffusion appears both to orbits starting close to simple as well as to double unstable periodic orbits. When this happens we observe plateaus in the $\overline{V}(t)$ variation.

\section{Conclusions}
 
In this paper we studied the structure of the phase space close to simple unstable and  double unstable periodic orbits in a triaxial 3D system 
representing a rotating disk galaxy. We have encountered different orbital behaviors and different diffusion rates from the neighborhood of the periodic orbit. The phenomenon of stickiness seems to be ubiquitous and does not depend on the kind of instability of the periodic orbit in the center of the region we study. 
Our results refer to orbits in the neighborhood of the families x1v2 (Skokos
et al 2002) and z-axis (Martinet \& de Zeeuw 1988) and can be 
summarized as follows: 
 
\begin{itemize}
\item The orbits starting in the neighborhood of a simple unstable periodic 
orbit form (a) double loop ribbons in the 4D space, (b) structures discernible in 
some projections after long integration times, (c) clouds of points around the
$U$ p.o. with mixing of colors in their 4D representations. The two first
kinds of orbital behavior appear close to the transition $U\rightarrow S$ of the x1 family at the point of
bifurcation of the x1v2 family   and indicate the presence of the phenomenon of stickiness in the dynamics of the system.
\item The asymptotic curves of the $U$ x1v2 p.o. close to the $U\rightarrow S$
  transition of the central family of periodic orbits x1 are composed of two
  parts. Part A (calculated for $c<0$ in (3)) winds around deformed rotational
  tori  belonging to orbits deviating from the x1 orbits in the $\pm z$ or $\pm \dot z$-direction. Part B (calculated for $c>0$ in (3)) winds around the  rotational tori of the symmetric x1v1 and x1v1$^{\prime}$ p.o. 
\item In both cases chaotic orbits are guided by the manifolds close to the
  regions occupied by the rotational tori associated with the stable periodic
  orbits existing in the region. These chaotic orbits are sticky and play a very important role, especially in Galactic Dynamics, since the stickiness time is in many cases of the order of a Hubble time or even larger. 
\item Double loops  and \rotatebox[origin=c]{90}{$\Theta \:$}-shaped structures surrounded by rings that are found in several 2D and 3D projections of the 4D spaces of section in 3D Hamiltonian systems are formed by orbits close to the asymptotic curves of simple unstable periodic orbits that surround the rotational tori of two nearby stable p.o. as in Fig.~\ref{x1v1v2}. Such configurations appear if the parent family (in our case the planar x1 family) has a $S\rightarrow U$ transition and a nearby $U\rightarrow S$ transition. At the first transition bifurcate two symmetric stable orbits (like x1v1 and x1v1$^\prime$) and at the second transition we have the bifurcation of two simple unstable families (in our case the families x1v2 and x1v2$^\prime$). Near the $U\rightarrow S$ transition of the parent family, the already bifurcated stable orbits are not far from the unstable orbit under study, thus the asymptotic curves of the latter families can surround their rotational tori.
Away from the transition points of the varying parameter (in our case the Jacobi constant) the shape of the asymptotic curves become complicated and the consequents of the nearby orbits form clouds in the 4D spaces of section.
\item Double instability in 3D rotating galactic disks appears for larger energies than simple instability in the families of the x1 tree. The usual case is after a $U\rightarrow DU$ transition and in general there are no rotational tori associated with stable families in the neighborhood of the $DU$ p.o.  Typically the consequents of the orbits form clouds of points in the 4D spaces of section.
\item In the case of a double unstable periodic orbit there are  two different pairs of 
eigendirections, starting at the periodic orbit. The 
curves of each pair start in opposite directions. Between the eigendirections 
an ``eigensurface'' is formed. However the shape of this surface is complicated and it is difficult to be visualized when the associated eigenvalues are not close to each other. 
\item In the case of z-axis $DU$ periodic orbits in slow pattern speed models,
  at low energies, we find that the eigenvalues corresponding to the unstable
  eigenvectors are close to each other. Because of that the speeds of
  deviation along the two eigendirections are similar. This allows us to
  visualize  the unstable eigensurface corresponding to the $DU$ periodic
  orbit. We find sticky chaotic orbits whose consequents  stay close to the unstable
  manifold for very long times (of the order of a Hubble time or more). We find these sticky orbits either by perturbing the initial conditions of the z-axis $DU$ p.o., or by perturbing orbits on rotational tori associated with the stable anomalous orbit (sao) and its symmetric family.
\item Among the orbits we studied we found those close to the double unstable
  orbits of the x1v2 family having the largest diffusion speed. The sticky
  chaotic orbits close to the bifurcating point of the simple unstable
  x1v2 orbit and close to the double unstable z-axis orbit we have examined,
  have comparable diffusion speeds. They are much slower than the diffusion
  speeds of the orbits in the neighborhood of x1v2 simple unstable periodic
  orbits away from its bifurcation point, or of the double unstable orbits of the same family that have
  eigenvalues of very different size.
\end{itemize}

These results will be used in a forthcoming paper to assess the role of chaotic orbits in shaping the vertical profiles of disk galaxies.

\vspace{2cm}
\textit{Acknowledgments} We thank Dr. C.~Efstathiou for comments on an earlier draft of this paper. \\

\section{References}
\begin{enumerate}
\item Arnold, V.I. [1988], 
{\em Geometrical Methods in the Theory of Ordinary Differential Equations} 
(2nd edition, Springer-Verlag NY).
\item Broucke, R. [1969]  
``Periodic Orbits in the Elliptic  Restricted  Three-Body   Problem'', 
\textit{NASA Tech. Rep.} 32-1360, 1-125.
\item Contopoulos, G. [2002] 
{\em Order and Chaos in Dynamical Astronomy} (Springer-Verlag, New York 
Berlin Heidelberg).
\item Contopoulos, G. [2009]  
``Ordered and Chaotic Orbits in Spiral Galaxies'' in {\em `Chaos in Astronomy}
, G. Contopoulos and P.A. Patsis (eds) (Springer-Verlag, Berlin, Heidelberg), 
  p.3-22.
\item  Contopoulos, G. \& Harsoula, M. [2008]  ``Stickiness in Chaos'',
 \textit{Int. J. Bif. Chaos} \textbf{18}, 2929-2949.
\item Contopoulos, G. \& Harsoula, M. [2010]  ``Stickiness effects in chaos'',
 \textit{Celest. Mech. Dyn. Astron.} \textbf{107}, 77-92.
\item Contopoulos, G. \& Magnenat, P. [1985]  ``Simple three-dimensional  
 periodic orbits in a galactic-type potential'' \textit{Celest. Mech.} 
\textbf{37}, 387-414.
\item Hadjidemetriou, J. [1975]  ``The stability of periodic orbits in the 
  three-body problem'', \textit{Celest. Mech.} \textbf{12}, 255-276.
\item Heisler, J., Merritt, D. \& Schwarzschild, M. [1982]  ``Retrograde closed
  orbits in a rotating triaxial potential'', \textit{ApJ} \textbf{258}, 490-498.
\item Katsanikas, M. \& Patsis, P.A. [2011]  ``The structure of invariant tori 
in a  3D galactic potential'', \textit{Int. J. Bif. Chaos} \textbf{21}, 
467-496. 
\item Katsanikas, M., Patsis, P.A. \& Contopoulos, G. [2011]  ``The structure 
and evolution of confined tori  near a Hamiltonian Hopf bifurcation'', 
\textit{Int. J. Bif. Chaos} \textbf{21}, 2321-2330. 
\item Magnenat, P. [1981] 
{\em \'{E}tudes num\'{e}riques sur le computement des orbites stellaires dans des modeles galactiques a 2 ou 3 degr\'{e}s de libert\'{e}} PhD Thesis, Observatoire de Geneve.
\item Magnenat, P. [1982] 
 ``Numerical study of periodic orbit properties in a   dynamical system  
with three  degrees of freedom'', \textit{Celest. Mech.}  \textbf{28}, 319-343.
\item Martinet, L. \&  de Zeeuw, T. [1988]
 ``Orbital stability in rotating triaxial stellar systems'', 
\textit{Astron. Astrophys.} \textbf{206}, 269-278.
\item Miyamoto, M.  \&  Nagai, R. [1975] 
 ``Three-dimensional models for the distribution of mass in galaxies'', 
\textit{Publ. Astron. Soc. Japan} \textbf{27}, 533-543.
\item Patsis, P. A. \&  Zachilas, L. [1990]  ``Complex instability of simple
  periodic orbits in a realistic two-component galactic potential'',\textit{
    Astron. Astrophys.} \textbf{227}, 37-48.
\item Patsis, P.A. \&  Zachilas, L. [1994]  ``Using Color and rotation for 
  visualizing  four-dimensional Poincar\'{e}  cross-sections: with 
  applications to the  orbital  behavior of a three-dimensional 
  Hamiltonian system'', \textit{Int. J. Bif. Chaos} \textbf{4}, 1399-1424.
\item Patsis, P.A., Skokos, Ch. \& Athanassoula, E. [2002] 
 ``Orbital dynamics of three-dimensional bars-III. Boxy/peanut edge-on 
profiles'', \textit{Mon Not. R. Astr. Soc.} \textbf{337}, 578-596.
\item Skokos, Ch. [2010]   ``The Lyapunov Characteristic Exponents and their 
   Computation'', \textit{Lect. Not. Phys.}  \textbf{790}, 63-135.
\item Skokos, Ch. [2001]  ``On the stability of periodic orbits of high 
    dimensional autonomous Hamiltonian systems'', \textit{Physica D} 
    \textbf{159}, 155-179.
\item Skokos,  Ch., Patsis, P.A. \&  Athanassoula, E. [2002] 
 ``Orbital dynamics  of three-dimensional bars-I. The backbone of 
three-dimensional bars. A fiducial  case'', \textit{Mon. Not. R. Astr. Soc.} 
\textbf{333}, 847-860.
\item Vrahatis, M.N., Isliker, H. \& Bountis, T.C. [1997] 
``Structure and breakdown of invariant tori in a 4-D mapping model of 
accelerator dynamics'', \textit{Int. J. Bif. Chaos.}  \textbf{7}, 2707-2722.

\end{enumerate}

\end{document}